\documentclass[prb,twocolumn,superscriptaddress,showpacs] {revtex4-1}
\usepackage{amsmath,amsfonts,amssymb,mathbbol,dsfont}
\usepackage{graphicx,psfrag,color}
\usepackage{dcolumn}
\usepackage{bm}
\usepackage{mathbbol, appendix}

\def\l{{\bm{l}}}
\def\d{{{\sf d}}}
\def\r{{\bm{r}}}

\def\i{{\bm{e_1}}}
\def\j{{\bm{e_2}}}

\begin{document}
\title{Arbitrary Dimensional Majorana Dualities and 
Architectures for Topological Matter}
\author{Zohar Nussinov}
\affiliation{Department of Physics, Washington University, St.
Louis, MO 63160, USA}
\author{Gerardo Ortiz}
\affiliation{Department of Physics, Indiana University, Bloomington,
IN 47405, USA}
\author{Emilio Cobanera}
\affiliation{Department of Physics, Indiana University, Bloomington,
IN 47405, USA}
 
\date{\today}

\begin{abstract}
Motivated by the prospect of attaining Majorana modes at the ends of
nanowires, we analyze interacting Majorana systems on general networks
and lattices  in an arbitrary number of dimensions, and derive various
universal spin duals. Such general complex Majorana architectures (other than those of 
simple square or other crystalline arrangements) might be of empirical relevance.
As these systems display low-dimensional symmetries, they are candidates for  realizing topological quantum
order. We prove that (a) these Majorana systems, (b) quantum Ising gauge
theories, and (c) transverse-field  Ising models with annealed bimodal
disorder are all dual to one another on general graphs. This leads to an
interesting connection between heavily disordered annealed Ising systems
and uniform  Ising theories with nearest-neighbor interactions. As any
Dirac fermion (including electronic) operator can be expressed  as a
linear combination of two Majorana fermion operators, our results
further lead to dualities  between interacting Dirac fermionic systems
on rather general lattices and graphs and corresponding spin systems.
The spin duals allow us to predict the feasibility of various standard
transitions as well as spin-glass type  behavior in {\it interacting}
Majorana fermion or electronic systems. Several  new systems that can be
simulated by arrays of Majorana wires are further introduced  and
investigated: (1) the {\it XXZ honeycomb compass} model (intermediate between
the classical Ising model on the honeycomb lattice and Kitaev's
honeycomb model), (2) a checkerboard lattice realization of the model 
of Xu and Moore for superconducting $(p+ip)$ arrays, and a (3) compass
type two-flavor Hubbard model with both pairing and hopping terms. By
the use of dualities, we show that all of these systems lie in the 3D
Ising universality class. We discuss how the existence  of topological
orders and bounds on autocorrelation times can be inferred by the use
of symmetries and also propose to engineer {\it quantum simulators} out of these 
Majorana networks.  
\end{abstract}

\pacs{05.30.-d, 03.67.Pp, 05.30.Pr, 11.15.-q}
\maketitle
 
\section{Introduction}
Majorana (contrary to Dirac) fermions are particles that constitute
their own  anti-particles. \cite{Majorana} Early quests for Majorana
fermions centered on neutrinos and fundamental issues in particle
physics that have yet to be fully settled. If neutrinos were Majorana
fermions then neutrinoless double $\beta$ decay would be possible and
thus  experimentally observed.  More recently, there has been a flurry
of activity in the study of Majorana fermions  in candidate condensed
matter realizations,  
\cite{Beenakker,wilczek,franz,Alicea,fu,carlo,kitaev-wire,sau,
oreg,Biswas,Lai,Ivanov,Green,gerardo,lfu} including lattice
\cite{Kitaev06,xulu,Terhal} and other\cite{kitaev-wire,sau}  systems 
inspired by  the prospect of topological quantum computing. 
\cite{kitaev03,Nayak} In the condensed matter arena,  Majorana fermions
are, of course, not fundamental particles but rather emerge  as
collective excitations of the basic electronic constituents. The systems
discussed in this work form a generalization of a model \cite{Terhal}
that largely builds and expands on ideas considered by Kitaev 
\cite{kitaev-wire,Kitaev06, kitaev03} including, notably,  the
feasibility of creating Majorana fermions at the endpoints of
nanowires.  \cite{lieb-shultz-mattis} A quadratic fermionic Hamiltonian
for electronic hopping along a wire in the presence of superconducting
pairing terms (induced by a proximity effect to bulk superconducting
grains on which the wire is placed) can be expressed as a Majorana 
Fermi bilinear that may admit free unpaired Majorana Fermi modes at the
wire endpoints. \cite{lieb-shultz-mattis} Kitaev's proposal entailed
$p$-wave superconductors. \cite{kitaev-wire}

More recent and detailed studies suggest simpler and more concrete ways
in which zero-energy Majorana modes might explicitly appear at the
endpoints  of nanowires placed close to (conventional $s$-wave)
superconductors. Some of the best  known proposals \cite{carlo,sau,oreg}
entail semiconductor nanowires  (e.g., InAs or InSb \cite{kouwenhoven}) 
with strong depolarizing Rashba spin-orbit coupling that are immersed in
a magnetic field that leads to a competing Zeeman effect. These wires
are to be placed close to superconductors in order  to trigger
superconducting pairing terms in the wire.  By employing the
Bogoliubov-de Gennes equation to study the band structure,  it was
readily seen how Majorana modes appear when the band gap vanishes. 
\cite{carlo,sau,oreg}  Along another route, it was predicted that
zero-energy Majorana fermions might appear at  an interface between a
superconductor and a ferromagnet.  \cite{fu,xulu} Majorana modes may
also appear in time-reversal invariant $s$-wave topological
superconductors.  \cite{gerardo}

If zero-energy Majorana fermions may indeed be {\it harvested} in these
or other ways \cite{carlo} then it will be natural to consider what
transpires in general networks made of such nanowires. The possible rich
architecture of structures constructed out of Majorana wires and/or
particular junctions  may allow for interesting collective phenomena as
well as long sought topological quantum computing applications. 
\cite{kitaev03,Nayak} Interestingly, as is well appreciated, the
braiding of (degenerate) Majorana fermions realizes a non-Abelian
unitary transformation that may prove useful in quantum computing
providing further impetus to this problem. In the current work, we
consider general questions  related to Majorana Fermi systems that may
be constructed from nanowire architectures. 

A central question regarding systems of Majorana fermions is concerned
with  viable {\it topological quantum orders} (TQOs). Disparate  (yet
inter-related) definitions of TQO appear in the literature.  One of the
most striking (and experimentally important) aspects of TQO is its 
immunity to local perturbations or, equivalently, its inaccessibility to
local probes at  both zero and finite temperatures. \cite{TQO} Some of
the best studied TQO systems are Quantum Hall fluids. 
\cite{Nayak} Several lattice models are also well known to  exhibit TQO,
including the spin $S=1/2$ models introduced by Kitaev. 
\cite{Kitaev06,kitaev03}  In the context  of the Majorana lattice
systems (and general networks) that we investigate here,  one currently
used approach for assessing the presence of TQO  \cite{Terhal} is
observing whether a fortuitous match occurs, in perturbation theory,
between (a) the studied nanowire systems with (b) Hamiltonians of
lattice systems known to exhibit TQO.  While such an analysis is highly
insightful, it may be hampered by the limited number  of lattice systems
(and  more general networks) that have already been established to
exhibit TQO.

In this work we suggest a different method for constructing Majorana
system architectures displaying TQO. This approach does not require us
to work towards an already examined lattice system  that is known to
exhibit TQO. Instead, our recipe invokes  direct consequences of quantum
invariances.  Symmetries can mandate and protect the  appearance of TQO
\cite{TQO}  via a generalization of Elitzur's theorem. \cite{BN,ads}
Specifically, whenever {\it $d$-dimensional gauge-like symmetries}
\cite{TQO} are present  (most importantly, discrete $d=1$ or continuous
$d=1,2$ symmetries),  finite temperature TQO may be mandated.
Zero-temperature TQO states protected by
symmetry-based selection rules can be further constructed.  A symmetry
is termed a $d$-dimensional gauge-like symmetry if it involves
operators/fields that reside in a $d$-dimensional volume. 
\cite{TQO,BN,ads} The use of symmetries offers a direct route for
establishing TQO that does not rely on particular known models as a
crutch for establishing its presence. 

\begin{figure}[h]
\centering
\includegraphics[angle=0,width=.8\columnwidth]{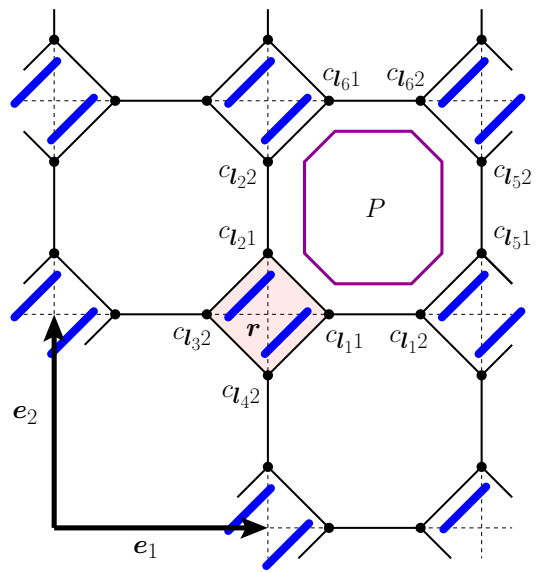}
\caption{A decorated square lattice (with unit vectors $\i$ and $\j$) in
which each site is replaced by a tilted square (representing a
superconducting grain at site $\r$). Two nanowires (solid blue diagonal
lines)  are placed on each grain.  The grains are coupled to each other
via  Josephson couplings. A local (gauge) symmetry operator of the model
is  $G_P=(i c_{\l_1 1} c_{\l_1 2})(i c_{\l_5 1} c_{\l_5 2})(i c_{\l_6
1}  c_{\l_6 2})(i c_{\l_2 1} c_{\l_2 2})$, where $P$ defines the minimal
closed  loop. See text.}
\label{l12}
\end{figure}

To illustrate the basic premise as it may be applied to architectures
with Majorana  fermions, we will advance and study a generalization of a
model introduced in  Ref. \onlinecite{Terhal} to describe a square
lattice array of Josephson-coupled nanowires on superconducting grains.
A schematic of the array studied in Ref. \onlinecite{Terhal} is
presented in Fig. \ref{l12}. As we will elaborate on in Section
\ref{systems},  our general-dimensional extension of this Hamiltonian is
given by Eqs. (\ref{Ham}),(\ref{star*}), and (\ref{d2}) with $c_{\l i},
(i=1,2)$ denoting Majorana operators  (satisfying the standard Majorana
algebra of Eq. (\ref{majorana_algebra})) associated with nanowire
endpoints. Within the generalized scheme, these nanowires are placed on
superconducting islands that occupy the vertices $\r$ of a general 
(even-coordinated) network, with links \(\l\) connecting the islands.
The ends of the nanowires are placed so that each link \(\l\)  connects
two Majorana fermions $c_{\l 1}, c_{\l 2}$ from different wires. Each
link carries an arbitrary but fixed orientation, just for the purpose of
labelling the Majoranas on it: As one traverses a link in the specified
direction, $c_{\l 1}$ comes before $c_{\l 2}$ (see Fig. \ref{l12}). 

For example, in Fig. \ref{l12}, two parallel nanowires are placed on
each superconducting grain. These grains are placed on the sites $\r$ of
a square lattice matrix. The two nanowires on each grain yield four
Majorana fermionic degrees of freedom, placed on the edges of the
oriented links of the lattice.  The Majorana fermions on different
superconducting grains, sharing a link, are coupled  to each other by
Josephson junctions. Prior to introducing the Josephson couplings, each 
grain is shunted to maintain a fixed superconducting phase and is
capacitively coupled to  a ground plate. Consequently, there are large
fluctuations in the electron number operator. However, the electron number parity is
conserved.  The sum of the two dominant effects: {\bf{(i)}} inter-grain
Josephson couplings and  {\bf{(ii)}} intra-grain  constraints on the
electron-number parity, complemented by exponentially small capacitive
energies, leads to a simple effective Hamiltonian.  The intra-grain
constraint on electron number (even/odd) parity is more dominant than 
inter-grain effects. The parity operator is ${\cal P}_{\r} =
(-1)^{n_{\r}}$ with $n_{\r}$  the total number of electrons on grain
$\r$. This electron number parity can be of paramount importance
in interacting Majorana systems.  \cite{lfu,xulu} In grains having two nanowires each, the electronic parity
operator is quartic in the  Majorana fermions; it is just the ordered
product of the four Majorana fermions at  the endpoints of the nanowires
on top of the grain at site $\r$, 
\begin{equation}
{\cal P}_{\r}=c_{\l_1 1}c_{\l_2 1} c_{\l_3 2} c_{\l_4 2},\ \ \ \ 
\r\in \l_1,\l_2,\l_3,\l_4
\end{equation} 
(we write \(\r\in \l\) to indicate that \(\r\) is one of the two
endpoints of \(\l\)). This gives rise to a term in the effective
Hamiltonian of the form  \cite{Terhal}
\begin{eqnarray}
H_{0} = - h \sum_{\r} {\cal{P}}_{\r},
\label{H00}
\end{eqnarray} 
with the sum taken over all grains, whose total number is $N_\r$. 
This term is augmented by Josephson
couplings across  inter-grain links $\l$,  leading to a Majorana Fermi
bilinear term involving the coupled pair of Majoranas $\{(c_{\l1},
c_{\l2})\}$,
\begin{eqnarray}
\label{H11}
H_{1} = -J\sum_{\l} ic_{\l1} c_{\l2}.
\end{eqnarray}
Fermionic parity effects are more dominant than Josephson coupling ($h
\gg J$) effects. Invoking perturbation theory, for small $(J/h)$, it was
found \cite{Terhal} that, to lowest non-trivial order, the  resultant
effective Hamiltonian was identical to that of Kitaev's toric code
model, \cite{kitaev03} thus establishing that such a system may support
TQO. Unfortunately, for $(J/h) \ll 1$,  spectral gap is small and the
system is more susceptible to thermal fluctuations and noise. A 
Jordan-Wigner transformation was  invoked \cite{Terhal} to illustrate
that these results survive for finite $(J/h)$. 

In this article,  we will outline a general procedure for the design  of
different architectures of nanowires on superconducting grains  that
support TQO. As alluded to above, our considerations will not be 
limited to the use of perturbation theory but will rather  rely on the
use of symmetries and exact generalized dualities associated with  these
{\it granular} and other systems defined on general networks. We will
further invoke a general framework for dualities that does not require 
the incorporation of known explicit representations of a spin in terms
of Majorana  fermions nor Jordan-Wigner transformations that have been
invoked  in earlier works. \cite{xulu,Terhal,Fradkin} The
bond-algebraic  approach, \cite{ads,NO,bond,orbital,bondprl,ADP,clock}
that we employ to study general {\it exact} dualities and
fermionization, \cite{bondprl,ADP}  allows for the derivation of earlier
known dualities as well as a plethora of many new others  for rather
general networks (or graphs) in arbitrary dimensions and boundary
conditions. It is important to note, as we will return to explicitly
later,  that as Dirac fermions can be expressed as a linear combination
of two Majorana fermions, our mappings lead to dualities between
standard (non-Majorana) fermionic systems and  spin systems on arbitrary
graphs in general dimensions. These afford non-trivial examples of
fermionization in more that one dimension. 

Among several exact dualities that we report here we note, in
particular, the following: 

\begin{itemize}
\item
A duality, in any dimension, between the Majorana fermion system
corresponding  to an arbitrary network of nanowires on superconducting
grains and quantum Ising gauge theories. 

\item
A gauge-reducing and emergent dualities \cite{ADP} in arbitrary number of dimensions
between granular Majorana Fermi systems on an arbitrary network and
transverse-field  Ising models with annealed exchange couplings. In two
dimensions, this duality, along with the first one listed above,
indicates that  an annealed average over a random exchange may leave the
system identical to a  uniform transverse-field Ising model. 
 
\item 
A further duality between a particular Majorana fermion architecture
and a nearest-neighbor quantum spin $S=1/2$ model which, in some sense,
is intermediate  between an Ising model on a honeycomb lattice and the
Kitaev honeycomb model.  \cite{Kitaev06} We term this system the ``XXZ
honeycomb compass model''. 
\end{itemize}

As one of the key issues that we wish to address concerns viable TQO, 
boundary conditions may be of paramount importance. Boundary conditions
are inherently related  to the character (and, on highly connected
systems, to the number) of  independent $d$-dimensional gauge-like
symmetries.  Imposing periodic or other boundary conditions on a system
can lead to vexing problems in traditional approaches to dualities and
fermionization.  By using bond-algebras, we can circumvent these
obstacles and construct exact dualities for  both infinite systems and
for finite systems endowed with arbitrary boundary conditions. Other
formidable barricades, such as the use of non-local string
transformations, can be overcome as well within the bond-algebraic
approach to dualities. \cite{ADP} The validity of any duality mapping
can, of course, be checked numerically by establishing that the spectra
of the two purported dual finite systems indeed coincide. The matching
of the spectra  serves as a definitive test since dualities are (up to
global redundancies) unitary transformations  \cite{ADP} that preserve
the spectrum of the system.
 
\begin{figure}[th]
\centering
\includegraphics[angle=0,width=.9\columnwidth]{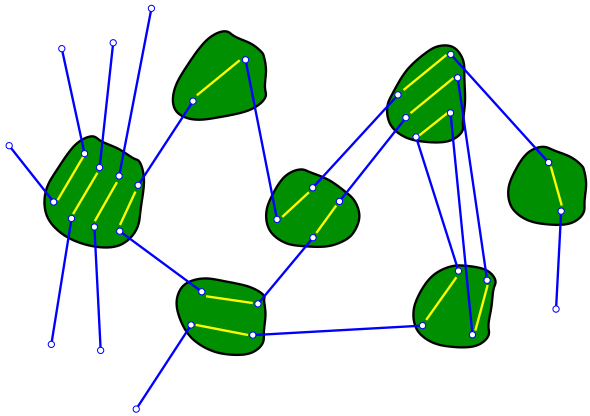}
\caption{A general network of superconducting grains with an even
coordination number  of each vertex. The local coordination number
$q_{\r}$ of any superconducting grain  centered about site $\r$ is equal
to the number of endpoints of all nanowires that are placed on that
superconducting grain. The  dominant Josephson tunneling paths between
inter-grain nanowire endpoints are highlighted by solid lines. Shown
here  is a two-dimensional projection of the network.}
\label{network}
\end{figure}

\section{Networks of Superconducting Grains and Nanowires}
\label{systems}

\begin{figure}[h]
\centering
\includegraphics[angle=0,width=.9\columnwidth]{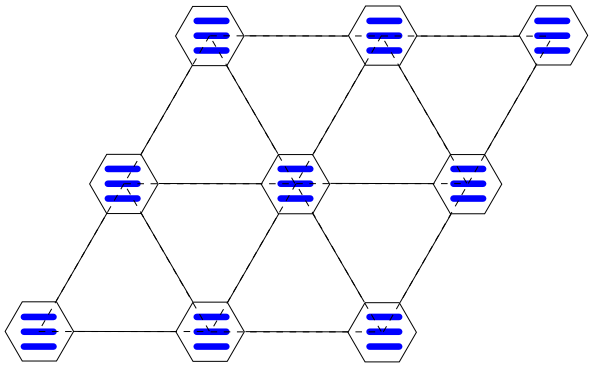}
\caption{A triangular network of superconducting grains (hexagons) on
each of which we place three nanowires.}
\label{l13}
\end{figure}

In the Introduction, we succinctly reviewed the effective Hamiltonian
for the square lattice array,\cite{Terhal} depicted in Fig. \ref{l12},
of Josephson-coupled  granular superconductors carrying each two
nanowires. This architecture serves as a  useful case of study. {\it
There is more to life, however,  than square lattice arrays} (although
we will return to these later on in this work).  We consider next rather
general architectures in which each node $\r$ (superconducting grain)
has an even number of nearest neighbors to which it is linked by
Josepshon coupling,  see Fig. \ref{network}. These general networks
include, of course, any two dimensional  lattice of even coordination,
e.g, those of Figs. \ref{l12}, and \ref{l13}, as special cases. 

The architectures that we consider are  realized by placing at each
vertex $\r$ of a graph-theoretical network a finite-size superconducting
grain. On each of these grains 
there are $z_{\r}$ nanowires. These nanowires provide  $2z_{\r}$
Majorana fermions, one for each wire's endpoint. Inter-grain Josephson
tunneling is represented by a link involving  Majoranas coming from
different wires on different islands. We place the nanowires on every
grain in the network so that each endpoint of a nanowire is near the
endpoint of another nanowire on a neighboring grain, to maximize
Josephson tunneling. Thus, the coordination number $q_{\r}$ of grain
$\r$ in these graphs is $q_{\r} = 2z_{\r}$.\cite{combinatorics} The
general situation is depicted in Fig. \ref{network}. 

The basic inter and intra-island interactions have different origins.  
For ease of reference, we reiterate these below  for arbitrary networks: 
\begin{itemize}
\item
there is a Josephson coupling $J_{\l}$ associated  with each inter-grain
link $\l$ of the  network connecting different superconducting grains,
and 
\item
an intra-grain charging energy $h_{\r}$ associated to each island 
at site \(\r\). 
\end{itemize} 

In a general, spatially non-uniform, network the spatial distribution of
couplings $J_{\l}$ and charging energies $h_{\r}$ need not be constant. 

The algebra of Majorana fermions is defined by the following relations:
\begin{eqnarray}
\{c_{\l i},c_{\l' i'}\}=2\delta_{\l,\l'}\delta_{i,i'},\ \ \ \ 
c_{\l i}^\dagger=c_{\l i}.
\label{majorana_algebra}
\end{eqnarray} 
With all of the above preliminaries in tow,\cite{irreps} we are now ready to present
the effective Hamiltonian for the systems under consideration, 
\begin{eqnarray}\label{Ham}
H_{\sf M} =-i\sum_{\l} J_{\l}c_{\l1}c_{\l2}-\sum_{\r} h_{\r}{\cal
P}_{\r},
\end{eqnarray}
where 
\begin{equation}\label{star*}
{\cal{P}}_{\r} \equiv i^{z_{\r_2}}\ c_{\l_1 i_1}  c_{\l_2 i_2} \cdots
c_{\l_{q_{\r}} i_{q_{\r}}}  , \ \r\in \l_1,\cdots,\l_{q_{\r}} ,
\end{equation}
is the product of all Majorana fermion operators  associated with the
superconducting grain at site \(\r\), ordered in some definite but
arbitrary fashion (differing orderings produce the same operator up to a
sign). \cite{order_explain}

The index $i_m$  can be either  $i_m=1 $ or $i_m=2$, depending on the
particular orientation that has been assigned to  the links in the
network. More precisely,  $i_m=1$ if \(\l_m\) points away from \(\r\),
and  $i_m=2$ if \(\l_m\) points into \(\r\).  The factor
\(i^{z_{\r_2}}\) is introduced to  render \({\cal P}_\r\) self-adjoint.
Since
\begin{equation}
(c_{\l_1 i_1} \cdots c_{\l_{q_{\r}} i_{q_{\r}}})^\dagger=
(-1)^{q_\r(q_\r-1)/2}c_{\l_1 i_1} \cdots c_{\l_{q_{\r}} i_{q_\r}}
\end{equation}
and \(q_\r=2z_\r\), we set the integer \(z_{\r_{2}}\) to be the number 
of nanowires counted modulo \(2\), 
\begin{equation}
z_{\r_{2}}=\left\{
\begin{array}{lcl}
0& \mbox{if}& z_{\r}\ \mbox{is even}\\
1& \mbox{if}& z_{\r} \ \mbox{is odd}
\end{array}
\right.  .
\label{d2}
\end{equation}

As we remarked earlier, the operators ${\cal{P}}_{\r}$ are related to the  operators $n_{\r}$
counting the total number of electrons on the grain $\r$ as 
\begin{equation}
{\cal P }_{\r}=(-1)^{n_\r},
\end{equation}
thus measuring the {\it parity} of the number of electrons at site
\(\r\). Hamiltonian (\ref{Ham}) constitutes an arbitrary dimensional
generalization of the sum of the two terms in Eqs. (\ref{H00},
\ref{H11}). In the following, we call  the operators
$\{ic_{\l1}c_{\l2}\}$ and $\{{\cal{P}}_\r \}$ the {\it bonds} of the
Hamiltonian  \(H_{\sf M}\).\cite{bondprl,ADP}

\section{Symmetries and Topological Quantum Order}
\label{sym_section}

For the particular case of the square lattice ($D=2$), the interacting
Majorana Hamiltonian \(H_{\sf M}\) with periodic (toroidal) boundary
conditions was found  to exhibit \(0\)-dimensional local,
\(d=1\)-dimensional gauge-like, and \(2\)-dimensional global  
symmetries. \cite{Terhal} These symmetries, inherently tied to TQO
\cite{TQO} and dimensional reduction,  \cite{TQO,BN,ads} also appear in
the more general network renditions of the granular system just
described in the previous section.  They are also manifest for the
interacting Majorana systems embedded in any spatial dimension $D \ge 2$
when different boundary conditions are imposed.\cite{topology} 

{\underline{Global Symmetry}:} \newline
The Hamiltonian \(H_{\sf M}\) of Eq. \eqref{Ham} displays a global 
symmetry \(Q\), given by the product of all the Majorana Fermion
operators in the system. We can write \(Q\) in terms of bonds as
\begin{equation}\label{gsymmm}
Q=\prod_\r {\cal P}_\r,
\end{equation}
since each Majorana is contributed by some island.  The order of the
bonds in \(Q\) is not an issue,  since 
\begin{equation}
[{\cal P}_\r, {\cal P}_{\r'}]=0,
\end{equation}
for any pair of sites \(\r,\r'\). The conserved charge \(Q\) represents
a \(\mathbb{Z}_2\) symmetry of the system,
\begin{equation}
Q^2=\mathds{1}.
\end{equation}

Beyond this global symmetry, the system of Eq. (\ref{Ham}) exhibits
independent symmetries that operate on finer, lower-dimensional regions
of the network. 
Of particular importance to TQO are $d=1$ and $d=0$-dimensional
symmetries,  and so we turn to these next.

{\underline{$d=1$ symmetries}:} \newline
The $d=1$ dimensional  symmetry operators of the Majorana system are
given by
\begin{eqnarray}
Q_{\ell}=\prod_{\l \in \ell} (i c_{\l1}c_{\l2}),\ \ \ \
(Q_{\ell})^2=\mathbb{1},
\label{d1}
\end{eqnarray}
where $\ell$ is a continuous contour, finite or infinite and  open or
closed depending on boundary conditions,
entirely composed of links. That these non-local operators are
symmetries  is readily seen once it is noted that ({\cal{a}}) each of
the terms (or {\em bonds}) in the summand of Eq. (\ref{Ham})  defining
\(H_{\sf M}\) involves products of an even number of Majorana fermions
and ({\cal{b}}) by the second of  Eqs. (\ref{majorana_algebra}),
effecting an even number of permutations of Majorana fermion operators
in a product incurs no sign change. For example, for a network of linear
dimension $L$ along a  Cartesian axis,  the contour $\ell$ spans
${\cal{O}}(L^{1})$ sites and is thus a $d=1$ dimensional object. This is
the origin of the name $d=1$ symmetries.  Some of these $d=1$ symmetries
may be related to (appear as products of) the local  symmetries
discussed next, depending on the topology enforced by boundary 
conditions. Some others are fundamental and cannot be expressed in terms
of those local symmetries.
 
{\underline{$d=0$ symmetries}:} \newline
For the models under consideration, local, also called gauge, $d=0$
symmetries  are associated with the elementary loops (or plaquettes) $P$
of  the wires, see Fig. \ref{l12} for an example. That is,  when
considering  the superconductors as point nodes, the  links $\l$ form a
network with minimal closed loops $P$. The associated local symmetries
are given by 
\begin{eqnarray}
G_{P} = \prod_{\l\in P} (i c_{\l1}c_{\l2}), \ \ \ \ G_P^2=\mathds{1}.
\label{p*}
\end{eqnarray}
Repeating the considerations of ({\cal{a}}) and ({\cal{b}}) above, we
see that, for any elementary plaquette \(P\), the product of Majorana
Fermi operators in Eq. (\ref{p*}) commutes with  \(H_{\sf M}\), since it
shares an even number (possibly zero) of Majorana fermions with any bond
in the Hamiltonian. By multiplying operators $G_{P}$ for a collection 
of plaquettes $P$ that, together, tile  a region bounded by the loop
$\Gamma$, it is readily seen that this product is also a symmetry, as in
standard theories with gauge symmetries. 


The symmetries above lead to non-trivial consequences:

{\bf{(A)}} By virtue of Elitzur's theorem \cite{Elitzur} and its $d>0$
generalizations \cite{TQO,BN,ads}  all non-vanishing correlators 
$\langle \prod_{\alpha \in S} c_{\alpha} \rangle$  with $S$ a set of
sites $\alpha$ must be invariant under all of the symmetries of  Eqs.
(\ref{d1},\ref{p*}). That is, \(d=0,1\)-gauge-like symmetries cannot be
spontaneously broken. As we alluded to earlier, one consequence of the
non-local symmetries such as the $d=1$ symmetries of  Eq. (\ref{d1}) is
the existence of TQO. \cite{TQO,topology}

{\bf{(B)}} Bounds on autocorrelation times. As a consequence of the
$d=1$ symmetries of Eq. (\ref{d1}), and the  aforementioned
generalization of Elitzur's theorem as it pertains to temporal
correlators, \cite{ads} the Majorana Fermi system will exhibit finite
autocorelation times regardless of the system size. Of course, for
various realizations of dynamics and geometry of the   disorder,
different explicit forms of the autocorrelation times can be found. For
instance,  by use of bond-algebras, Kitaev's toric code model is
identical to that of a classical square plaquette model as in Ref.
\onlinecite{auto}. Similarly, Kitaev's toric code model \cite{kitaev03}
can be mapped onto two uncoupled one-dimensional Ising chains. 
\cite{bond,NO,TQO}  Different realizations of the dynamics can lead to
different explicit forms of $\tau$ in both cases,  however, finite
autocorrelation times are found in all cases (as they must be).
Similarly, more general than the exact bond algebraic mapping and
dimensional reductions that we find here, by virtue of $d=1$ symmetries
of Eq. (\ref{d1}), autocorrelation functions involving  Majorana
fermions on a line $\ell$ must be bounded by corresponding ones in a
$d=1$ dimensional system. \cite{ads} 

\section{
Dualities and spin realizations of arbitrary Majorana Architectures}
\label{main}

In this Section, we provide two spin duals to the interacting Majorana
system described by the effective Hamiltonian \(H_{\sf M}\)  of Eq.
(\ref{Ham}) on arbitrary lattices/networks. This applies to finite  or
infinite systems and for arbitrary boundary conditions. These two dual
systems are (1) quantum Ising gauge theories for $D=2$ systems, and more
general spin gauge theories in higher-dimensions, and (2) a family of
transverse-field Ising models with annealed disorder in the exchange
couplings (each model representing a single gauge sector of \(H_{\sf
M}\)). The  dualities will be established in the framework of the theory
of bond algebras of interactions,  \cite{bondprl,ADP} as it applies to the
study of general dualities between many-body Hamiltonians. 
The general bond algebraic method relies on a comparison of the 
algebras, in the respective two dual model, that are generated by the corresponding
local interaction terms (or {\it bonds}) in these theories. \cite{ads,NO,bond,orbital,bondprl,ADP,clock}  
For the problem at hand, the Hamiltonian \(H_{\sf M}\) is built as the sum of
two sets of Hermitian bonds
\begin{equation}\label{mbonds}
ic_{\l1}c_{\l2},  \ {\cal P}_\r ,
\end{equation}
where \(\l\) and \(\r\) are links and sites of the network supporting 
\(H_{\sf M}\) (\({\cal P}_\r\) was defined in Eq. \eqref{star*}). In
this paper, we will only consider the bond algebra \({\cal A}_{\sf M}\)
generated by these bonds. We can then obtain dual representations of
\(H_{\sf M}\) by looking for alternative local representations of 
\({\cal A}_{\sf M}\). But first we have to  characterize ${\cal A}_{\sf
M}$  in terms of relations. 

The problem of characterizing a bond algebra of interactions is
simplified by several features brought about by physical considerations
of locality.  The first consequence of locality is that interactions are
{\it sparse},  meaning that each bond in any local Hamiltonian
commutes with most other bonds and is involved in only a small number of relations
(or constraints) that link individual bonds to one another. Hence the number of non-trivial
relations per bond is small.  The second consequence is that relations
in a bond algebra can be classified into  intensive and extensive, and
most relations are intensive.  We call a relation {\it intensive} if the
number of bonds it involves is {\it independent of the size of the
system},  and extensive if the numbers of bonds it involves scales with
the size of the system. Since extensive relations  could potentially
lead to unphysical non-local behavior,  they are typically few in number
and may reflect the topology of the system regulated by the boundary
conditions, as we will illustrate repeatedly in this paper. 
As there are $(2z_\r)$ Majorana modes (or, equivalently, $z_\r$ 
fermionic modes) per grain, 
the Majorana theory of Eq. (\ref{Ham}) and the algebraic relations 
listed above are defined on a Hilbert space of dimension 
$\dim {\cal{H}}_{\sf M} = 2^{ z_{\r} N_{\r}}$

\begin{figure}[h]
\includegraphics[angle=0, width=.22\textwidth]{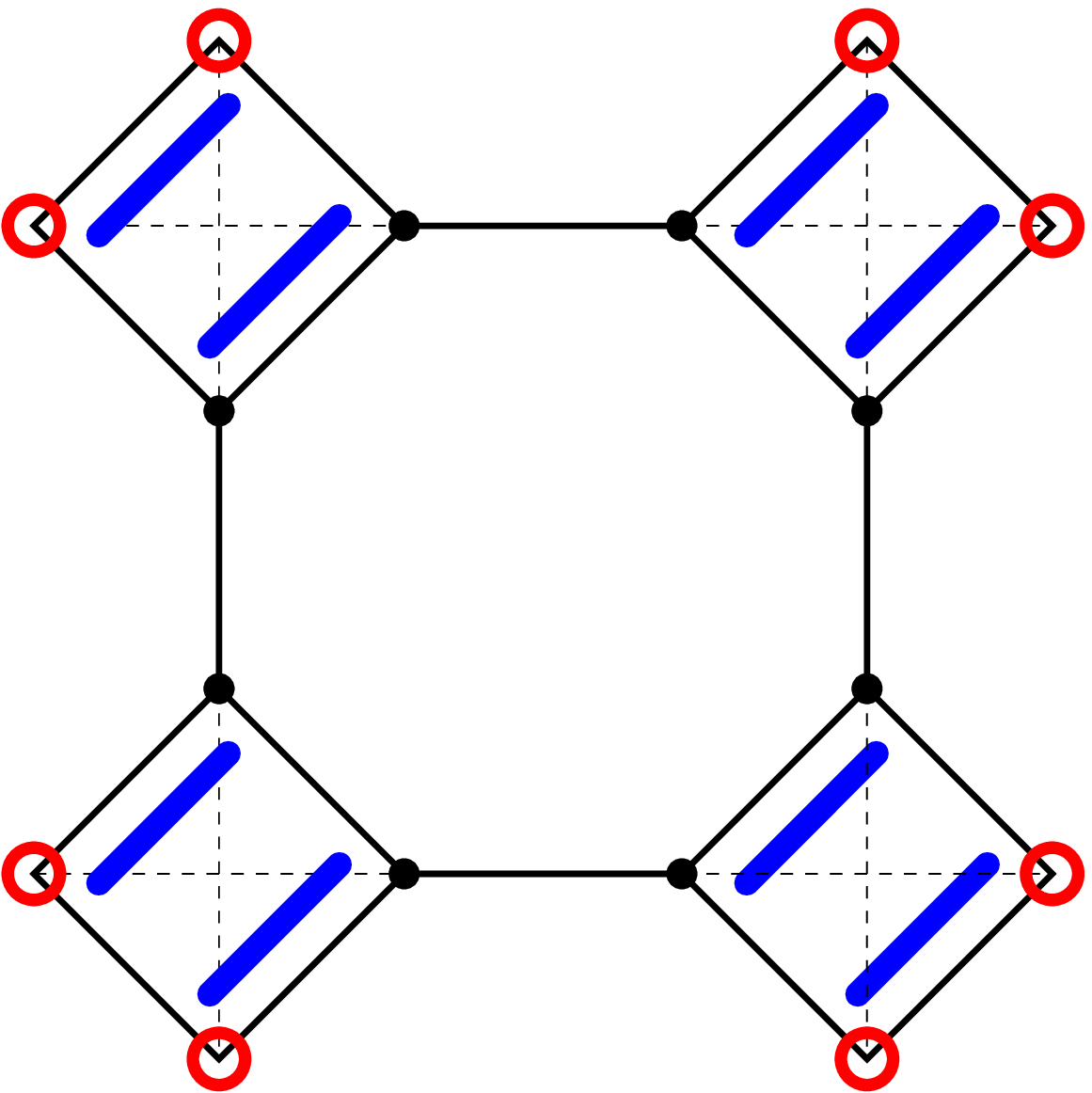} \ \ 
\ \ \ \ \includegraphics[angle=0, width=.22\textwidth]{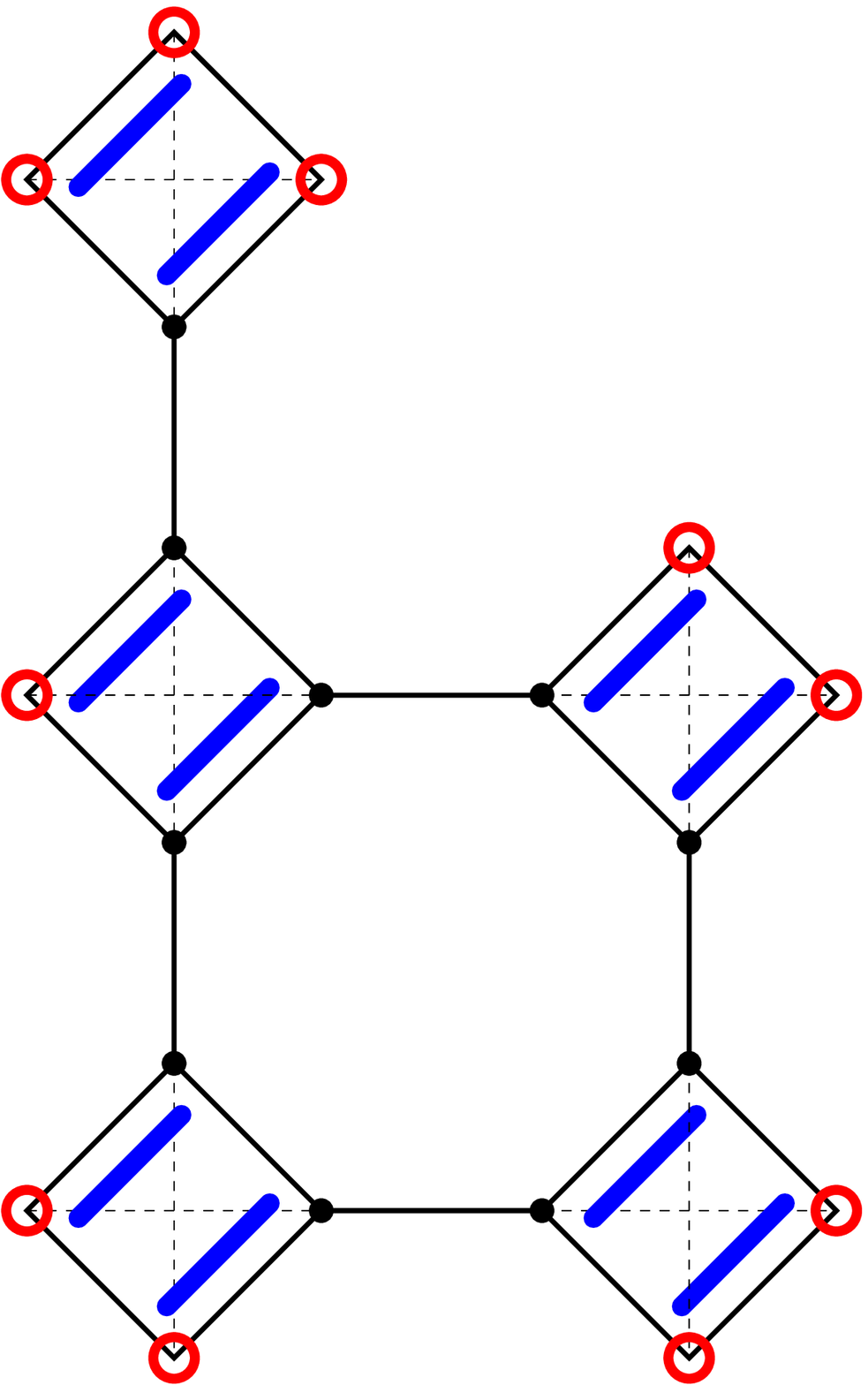}
\caption{Two architectures with open boundary conditions. In either
case, the operator \(B\) of Eq. \eqref{B} is the product of all the
uncoupled Majoranas on the boundary indicated by open circles, but
\([H_{\sf M}, B]=0\) only for the system shown in the panel on the
left.}
\label{bs}
\end{figure}

Next, we characterize the bond algebra ${\cal A}_{\sf M}$ as the first
step toward the construction of its spin duals. The intensive relations
are:
\begin{enumerate}
\item for any $\r$ and $\l$
\begin{equation}
(ic_{\l1}c_{\l2})^2=\mathds{1}=({\cal P}_\r)^2,
\end{equation}
\item
for \(\r,\r'\in \l\),
\begin{equation}
\{{\cal P}_\r,ic_{\l1}c_{\l2}\}=0=\{{\cal P}_{\r'},ic_{\l1}c_{\l2}\},
\end{equation}
\item
for \(\r\in \l_i,\ i=1,2,\cdots,q_\r\),
\begin{equation}
\{{\cal P}_\r,ic_{\l_i1}c_{\l_i2}\}=0.
\end{equation}
\end{enumerate}
Thus in the bulk, or everywhere for periodic boundary conditions, each
island anticommutes with four \(q_\r\) (the coordination of \(\r\))
links, and each link anticommutes with two
islands. The presence or absence of  extensive relations depends on the
boundary conditions. For  periodic (toroidal) or other closed boundary
conditions (e.g., spherical), we have one extensive relation 
\begin{equation}\label{ext_r}
\prod_\r {\cal P}_\r=
\alpha \prod_\l (ic_{\l1}c_{\l2}) , \ \ \ \ \alpha=\pm 1,
\end{equation}
since each Majorana Fermion operator appears exactly once both on the
left and right-hand side of this equation, but not necessarily in the
same order. The constant \(\alpha\) adjusts for the potentially
different orderings, and the overall powers of \(i\) on each side of the
equation. Notice that \(\prod_\r {\cal P}_\r=Q\) is the global
\(\mathds{Z}_2\) symmetry operator. In contrast, for open or semi-open
(e.g., cylindrical) boundary conditions, the islands on the free
boundary have Majorana fermions that are not matched  by links (that is,
that do not interact with Majoranas on other islands). Hence the product
\begin{equation}\label{B}
\left(\prod_\r {\cal P}_\r\right)\left(\prod_\l (ic_{\l1}c_{\l2})\right)=B
\end{equation}
reduces to the product \(B\) of these Majoranas on the free boundary.
The operator \(B\) may or may not commute with the Hamiltonian, 
depending on the details of the architecture at the boundary, see Fig.
\ref{bs}, but either way Eq. \eqref{B} does not represent an extensive
relation in the bond algebra (rather it just states how to write a
particular operator as a product of bonds). If \([H_{\sf M}, B]=0\),
\(B\)  represents a \(\mathds{Z}_2\) boundary symmetry independent of
the  local symmetries. 

\subsection{Duality to quantum Ising gauge theories}
\label{tdn}

In this Section we describe a duality relating the Hamiltonian \(H_{\sf
M}\) to a system of \(S=1/2\) spins. The spin degrees of freedom are
placed on the (center of the) {\it links} of a network identical to the
one  associated to \(H_{\sf M}\), and are described by Pauli matrices
\(\sigma^x_\l,\sigma^y_\l, \sigma^z_\l\).  The goal is to introduce
interactions among these spins that satisfy the same algebraic relations
as the bonds of \(H_{\sf M}\). Let us introduce the Hermitian spin bond
\begin{equation}
{\widetilde{\cal{P}}}_\r=\prod_{\{\l|\r\in \l\}} \sigma^z_\l.
\end{equation}
For example, for the special case of the square lattice discussed in the
Introduction,
\begin{equation}
{\widetilde{\cal{P}}}_\r=\sigma^z_{\l_1}\sigma^z_{\l_2}\sigma^z_{\l_3}
\sigma^z_{\l_4}, \ \ \ \ \r\in \l_1, \l_2, \l_3, \l_4. 
\end{equation} 

The set of spin bonds
\begin{equation}
\sigma^x_\l,\ \ \ \ \widetilde{\cal P}_\r,
\end{equation}
satisfy the following intensive relations
\begin{enumerate}
\item for any $\r$ and $\l$
\begin{equation}
(\sigma^x_\l)^2=\mathds{1}=(\widetilde{\cal P}_\r)^2,
\end{equation}
\item
for \(\r,\r'\in \l\),
\begin{equation}
\{\widetilde{\cal P}_\r, \sigma^x_\l\}=0=\{\widetilde{\cal
P}_{\r'},\sigma^x_\l \},
\end{equation}
\item
for \(\r\in \l_i,\ i=1,2,\cdots,q_\r\),
\begin{equation}
\{\widetilde{\cal P}_\r,\sigma^x_{\l_i}\}=0,
\end{equation}
\end{enumerate}
everywhere for closed boundary conditions, and everywhere in the bulk
for open or semi-open boundary conditions. These relations are identical
to the intensive relations for the bonds of \(H_{\sf M}\). In the Ising gauge theory, 
the bond algebraic relations listed above are defined on a space of size
$2^{z_{r} N_{\r}}$. (That this is so can be easily seen by noting that there 
are $N_{\l}=z_{\r} N_{\r}$ links each endowed with a spin $S=1/2$ degree of freedom 
 $\sigma^z_\l$.) As it so happens, this Hilbert space dimension 
 is identical to that of the Majorana system of $H_{\sf M}$. 
Putting all of the pieces together, we see that the
spin Hamiltonian
\begin{eqnarray}\label{QIG}
H_{\sf QIG}= -  \sum_{\l}  J_{\l} \sigma^x_\l-\sum_{\r}
h_{\r}\widetilde{\cal P}_\r
\end{eqnarray}
is {\it unitarily equivalent} to \(H_{\sf M}\), 
provided the extensive relations are matched as well. For open or
semi-open boundary conditions, the same follows provided that the intensive relations on the
boundary also properly match.  In the following, we focus on
periodic boundary conditions (of theoretical  interest in connection to
TQO), and leave the discussion of open boundary conditions (of interest
for potential experimental realizations of these systems) to the
Appendix \ref{appB}.

As just explained, the mapping of bonds
\begin{equation}\label{ppcc}
ic_{\l1}c_{\l2}\mapsto \sigma^x_\l,\ \ \ \
{\cal P}_\r \mapsto \widetilde{\cal P}_\r
\end{equation}
preserves the intensive algebraic relations. In particular it maps the
local symmetries of Eq. \eqref{p*} to local symmetries of \(H_{\sf
QIG}\),
\begin{equation}
G_P\equiv\prod_{\l\in P} (i c_{\l1}c_{\l2})\mapsto 
\prod_{\l\in P}\sigma^x_\l \equiv G_{S,P}.
\end{equation}
To assess the  effect it has on the extensive relation of Eq.
\eqref{ext_r} (and the global symmetry), notice that (for periodic
boundary conditions)
\begin{equation}
\prod_\l (ic_{\l1}c_{\l2})\mapsto \prod_\l \sigma^x_\l\equiv Q_S
\end{equation}
with \([Q_S,H_{\sf QIG}]\) a global symmetry of \(H_{\sf QIG}\),  and 
\begin{equation}
\prod_\r {\cal P}_\r \mapsto \prod_\r \widetilde{\cal P}_\r= \mathds{1}.
\label{rpr}
\end{equation}
It follows that, as it stands, the mapping of bonds of Eq.
\eqref{ppcc} is a correspondence, but not an {\it isomorphism} of
bond algebras. The simplest way to convert it into an isomorphism is to
modify one and only one of the bonds \(\widetilde{\cal P}_\r\) of the
spin model at some arbitrary site \(\r_0\), so that 
\begin{equation}
\widetilde{\cal P}_{\r_0}\equiv \alpha Q_S\prod_{\{\l|\r_0\in \l\}}
\sigma^z_\l
\end{equation}
(\(\alpha\) was defined in Eq. \eqref{ext_r})  while for any other site
\(\r\neq \r_0\), \(\widetilde{\cal P}_{\r}\) remains unchanged. The
introduction of this modified bond does not change the  intensive
relations, since \(Q_S\) commutes with every bond (original or
modified). Moreover,
\begin{eqnarray}
\prod_\r {\cal P}_\r \mapsto   \widetilde{\cal P}_{\r_0}\prod_{\r\neq
\r_0}\widetilde{\cal P}_\r=\alpha Q_S
\end{eqnarray}
and the extensive relation of Eq. \eqref{ext_r} is now, with the
modified definition of \(\widetilde{\cal P}_{\r_0}\), preserved, since
(\(\alpha^2=1\))
\begin{equation}
\prod_\l\sigma^x_\l=\alpha \widetilde{\cal P}_{\r_0}\prod_{\r\neq
\r_0}\widetilde{\cal P}_\r .
\end{equation}     
Hence there is a unitary transformation \(\mathcal{U}_\d\) such that 
\begin{equation}\label{MQIG}
\mathcal{U}_\d H_{\sf M} \mathcal{U}_\d^\dagger=H_{\sf QIG}, 
\end{equation}
with  \(H_{\sf QIG}\) containing the single modified bond
\(\widetilde{\cal P}_{\r_0}\). 

In the duality between the systems of Eqs. (\ref{Ham}, \ref{QIG}), the
dimensions of  the their Hilbert spaces  are identical. Since  we count
two Majorana modes (or, equivalently, one Fermionic mode) per link,  the
Hamiltonian \(H_{\sf M}\) is defined on a Hilbert space of dimension 
$\dim {\cal{H}}_{\sf M} = 2^{N_{\l}}$,  with \(N_{\l}\) denoting the
total number of links in the network. On the other hand, the spin system
has  one spin \(S=1/2\) degree of freedom per link, hence the dimension
of the Hilbert space on which $H_{\sf QIG}$ is defined is also
\(2^{N_{\l}}\).  Notice that the need to introduce the  modified bond
\(\widetilde{\cal P}_{\r_0}\) in the dual spin theory is irrelevant from
the point  of exploiting the duality to study the ground-state
properties of \(H_{\sf M}\) (or {\it viceversa}, to study the
ground-state properties of  \(H_{\sf QIG}\)), since for finite systems
the ground state \(|\Omega\rangle\) must satisfy \( 
Q_S|\Omega\rangle=|\Omega\rangle\). The ease with which we established
the duality between Majorana systems and QIG systems for general
lattices and networks illustrates how efficient the  bond-algebraic
construct is.

\begin{figure}[h]
\includegraphics[angle=0, width=.8\columnwidth]{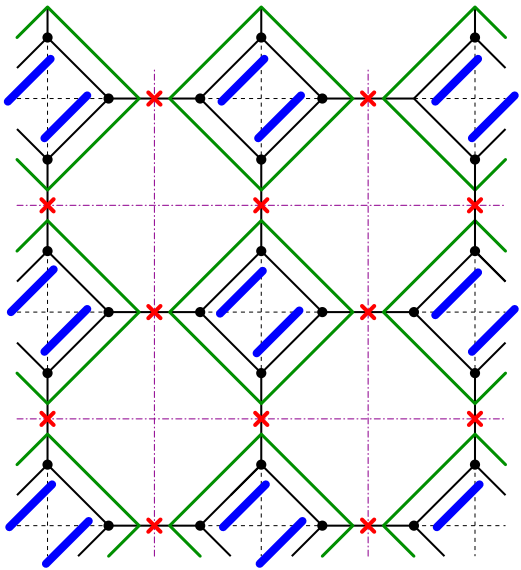}
\caption{Duality to a $D=2$   \(\mathds{Z}_2\) quantum Ising
gauge theory, where spins are represented as crosses. Hamiltonian 
\(H_{\sf M}\) of Eq. \eqref{ferm_wegner} 
represents a particular fermionization of the $H_{\sf QIG}$ gauge theory.}
\label{z2gf}
\end{figure}

The duality just described is extremely general, valid in particular for
any number of space dimensions $D$. In the following we  describe
explicitly one particularly important special instance,  that of $D=2$.
On a square lattice, the Hamiltonian \(H_{\sf M}\)   simplifies to 
\begin{equation}\label{ferm_wegner}
H_{\sf M}=-\sum_\l J_\l(ic_{\l1}c_{\l2})-\sum_\r h_\r c_{\l_11}c_{\l_2 1}
c_{\l_32}c_{\l_42},
\end{equation}
where \(\l_1,\l_2,\l_3,\l_4\) are shown in Fig. \ref{l12}. This 
generalizes the Hamiltonian considered in Ref. \onlinecite{Terhal} only
in that  inhomogeneous couplings are allowed. The dual spin (finite-size)
system is ($\r_0 \in \l_{0,1},\l_{0,2}\l_{0,3}\l_{0,4}$)
\begin{widetext}
\begin{equation}
H_{\sf QIG}=-h_{\r_0}\alpha Q_S\sigma^z_{\l_{0,1}}\sigma^z_{\l_{0,2}}
\sigma^z_{\l_{0,3}}\sigma^z_{\l_{0,4}}-\sum_{\r\neq\r_0} h_\r 
\sigma^z_{\l_1}\sigma^z_{\l_2}
\sigma^z_{\l_3}\sigma^z_{\l_4}-\sum_\l J_\l\sigma^x_\l
\end{equation}
\end{widetext}
that we recognize as the standard, $D=2$, \(\mathds{Z}_2\) Ising gauge
theory, \cite{fradkin_susskind} up to the modified bond at \(\r_0\)
(\(Q_S=\prod_\l\sigma^x_\l\) and \(\alpha\) is determined according to 
Eq. \eqref{ext_r}), see Fig. \ref{z2gf}. 

\begin{figure}[h]
\includegraphics[angle=0, width=.8\columnwidth]{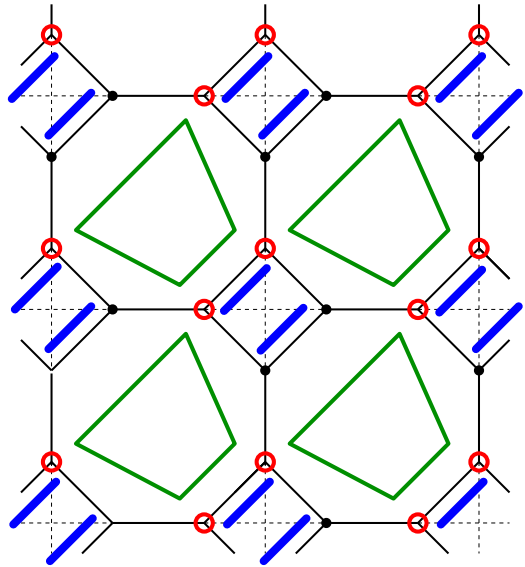}
\caption{Jordan-Wigner fermionization of the   \(\mathds{Z}_2\) Ising
gauge theory realizes a theory of Majorana fermions \(H_{\sf FSS}\),
with two-body interactions between Majoranas (shown as trapezoids)  on
three different islands. Notice that, unlike the intra-island two-body
interactions of \(H_{\sf M}\), two neighboring  two-body interactions
\(H_{\sf FSS}\) share a Majorana operator. }
\label{fss}
\end{figure}

Hence, we may regard the Hamiltonian of  Eq.
\eqref{ferm_wegner} as an exact fermionization of the \(\mathds{Z}_2\)
Ising gauge theory with periodic boundary conditions (and one modified
bond). It is interesting to compare this fermionization with a slightly
different  one  \cite{Fradkin} that exploits the Jordan-Wigner
transformation in the limit of infinite size. This approach yields the
Majorana  Hamiltonian \cite{Fradkin} (in our notation)
\begin{equation}
H_{\sf FSS}=-\sum_\l J_\l(ic_{\l1}c_{\l2})-\sum_\r h_\r c_{\l_12}c_{\l_5
1} c_{\l_62}c_{\l_21},
\end{equation}
where \(\l_1,\l_2,\l_5,\l_6\) are shown in Fig. \ref{l12}. {\it The
two-body  interaction \(c_{\l_12}c_{\l_5 1} c_{\l_62}c_{\l_21}\) is 
different  than the two-body interaction in \(H_{\sf M}\), since it
involves three different islands}, see Fig. \ref{fss}.  Hence,
disregarding boundary conditions, we see that the \(\mathds{Z}_2\)
quantum Ising gauge theory admits rather different but equivalent 
fermionizations. As  expected, the bonds  in \(H_{\sf FSS}\) satisfy
intensive relations identical to those already discussed for \(H_{\sf
M}\) and  \(H_{\sf QIG}\).

Thus far, we focused on periodic boundary conditions. We now 
remark on other boundary conditions.
When antiperiodic boundary conditions are imposed in a network with an 
an outer perimeter that includes twice an odd 
number of links, the right-hand side of Eq. (\ref{rpr}) is replaced by 
$- \mathds{1}$.  
The union of both cases (periodic and antiperiodic) for
a system having a twice odd perimeter spans all possible values of the product
$\prod_{\r} \widetilde{{\cal{P}}}_{\r}$.
Thus, for these systems in the case of periodic boundary conditions, the spectrum of 
the Majorana system can be mapped to the union of levels found for the {\sf QIG} systems
for both periodic or antiperiodic boundary conditions. In terms of the corresponding 
partition functions, we have that
\begin{eqnarray}
{\cal{Z}}_{\sf M, \, periodic} =  {\cal{Z}}_{\sf QIG, \, periodic} + 
{\cal{Z}}_{\sf QIG, \, antiperiodic}.
\end{eqnarray}

\subsection{Duality to annealed transverse-field Ising models}
\label{ttdn}

We next derive, in a similar spirit, a duality between the general
architecture Majorana system \(H_{\sf M}\) and annealed 
transverse-field Ising models. The number of annealed disorder variables 
in these systems (along with the number of sites $N_{\r}$) determines the size of
the Hilbert space on which the Ising models are defined. With an eye towards things to 
come, we note (as we will re-iterate later on) that the duality that we will 
derive in this Section will furnish an example in which
the Hilbert space dimensions of two dual systems need not be identical to one
another. Generally, dualities are unitary transformations between two theories up
to trivial gauge redundancies that do not preserve the Hilbert space dimension. \cite{ADP} 
That is, dualities are isometries.  

To define the annealed transverse field Ising systems, we place an \(S=1/2\)
spin on each site \(\r\), \(\sigma^x_\r, \sigma^y_\r,\sigma^z_\r\), of
the network associated to \(H_{\sf M}\), and a {\it classical}
annealed disorder variable $\eta_{\l} = \pm 1$ on each link \(\l\).
Then we can introduce the set of Hermitian  spin bonds
\begin{equation}\label{ispinb}
\sigma^x_{\r},\ \ \ \ \ \ \ \ \eta_\l\sigma^z_{\r}\sigma^z_{{\r'}},\ \ \r,\r'\in \l.
\end{equation} 
If we specialize  to periodic boundary conditions,  these bonds satisfy
a set of intensive relations identical to the ones discussed in the two
previous Sections, together with {\it one} new relation absent  before
and listed last below:
\begin{enumerate}
\item for any $\r$ and $\l$
\begin{equation}
(\sigma^x_{\r} )^2=\mathds{1}=(\eta_\l\sigma^z_{\r}\sigma^z_{{\r'}})^2,
\end{equation}
\item for $\r, \r' \in \l$
\begin{equation}
\{\sigma^x_{\r}, \eta_\l\sigma^z_{\r}\sigma^z_{\r'}\}=0
=\{\sigma^x_{\r'},\eta_\l\sigma^z_{\r}\sigma^z_{\r'}\},
\end{equation}
\item
for \(\r \in \l_i,\ i=1,2,\cdots, q_\r\),
\begin{equation}
\{\sigma^x_{\r},\eta_{\l_i}\sigma^z_{\r}\sigma^z_{\r_i'} \}=0,\ \ \ \ 
\r\neq \r_i'\in\l_i,
\end{equation}
\item 
for any elementary loop \(P\) in the network,
\begin{equation}\label{loopc}
\prod_{\l\in P} (\eta_\l\sigma^z_{\r}\sigma^z_{\r'}) =
\mathds{1}\prod_{\l\in P}\eta_\l.
\end{equation}
\end{enumerate}
The constraint of Eq. \eqref{loopc} holds true for any closed loop. For
this reason, and others related to TQO, it is  important to clarify the
meaning of {\it elementary loop}. 

Loops in the network that share some links can be joined along those
links to obtain another loop or sum of disjoint loops. This means that
the set of all loops has a minimal set of generators from which we can
obtain any loop or systems of loops by the joining operation just
described. We call the  loops in an arbitrary but fixed minimal
generating set {\it elementary loops}. In this way, we obtain a minimal
description of the constraints embodied in Eq. \eqref{loopc}. It is not
obvious a priori whether one should classify  these constraints (that
is, relations) as intensive or extensive. This depends on the topology
of the system. If the system is simply connected, every loop is
contractible to some trivial minimal (that is, of minimal length) loop,
and hence we can choose minimal loops as elementary loops.  These loops
afford an intensive characterization of the constraints embodied in  Eq.
\eqref{loopc}. If on the other hand the system is not simply connected,
as for periodic boundary conditions, the generating set of elementary
loops will include non-contractible loops, and the length of  some of
these non-contractible loops may scale with the size of  the system. 
Consider, for example,  the spin bonds of Eq.
\eqref{ispinb} on a planar network on the torus and on a punctured
infinite plane. Both networks fail to be simply connected,  but only the
torus forces some of the constraint of Eq. \eqref{loopc} to be
extensive, because its two non-contractible loops must scale with the
size of the system.   

For periodic boundary conditions, 
there is one extensive relation satisfied by the bonds of Eq. \eqref{ispinb},
\begin{equation}
\prod_\l (\eta_\l\sigma^z_\r\sigma^z_{\r'})=\eta\mathds{1},
\end{equation} 
with 
\begin{equation}
\eta\equiv\prod_\l \eta_\l,\ \ \ \ \eta=\pm 1,
\end{equation}
that may or may not be independent of the relations of Eq. \eqref{loopc},
depending on the details of the network. In the following, we will treat
it as an independent relation, since it does not affect our results if it turns
out to be dependent.
\begin{figure}[h]
\includegraphics[angle=0, width=.7\columnwidth]{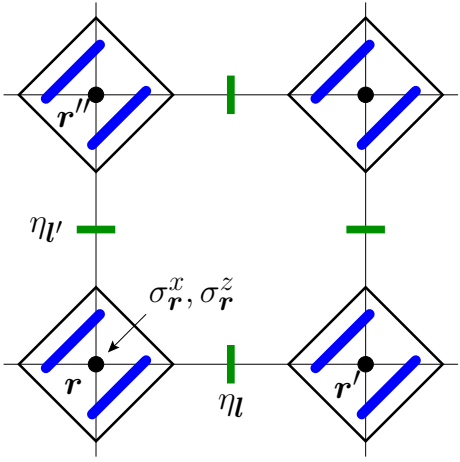}
\caption{Duality to an annealed transverse-field Ising model, in the 
particular $D=2$ case. Spins $S=1/2$ are located at 
the vertices $\r$ of the square lattice and classical $\mathbb{Z}_2$ 
fields $\eta_\l$ at the links $\l$ (indicated by a dash). }
\label{tfim2}
\end{figure} 

It follows that the mapping of bond algebras 
\begin{equation}
ic_{\l 1}c_{\l2}\mapsto \eta_\l\sigma^z_\r\sigma^z_{\r'}, 
\ \ \ \ {\cal P}_\r \mapsto \sigma^x_\r,
\end{equation}
preserves every local anticommutation relation. Hence 
the Hamiltonian theory
\begin{eqnarray}\label{hspin}
H_{\sf AI}\{\eta_{\l}\}= -\sum_{\l} J_{\l} (\eta_{\l}\sigma^z_\r
\sigma^z_{\r'})-\sum_{\r} h_\r \sigma^x_\r,
\end{eqnarray} 
obtained from applying this mapping to \(H_{\sf M}\)
will be shown to be dual to \(H_{\sf M}\) (see Fig. \ref{tfim2}). The 
Hilbert space on which the theory of Eq. (\ref{hspin}) is defined
is of size $\dim {\cal{H}}_{\sf AI} = 2^{N_{\r} + N_{\eta}}$ where  $N_{\r}$ is the
number of superconducting grains and $N_{\eta}$ the 
total number of $\eta_\l$ fields.

The proposed duality 
raises an immediate question: What are the features of \(H_{\sf M}\) that 
determine or at least constrain the classical fields \(\eta_\l\)? As 
we will see, the answer lies in the local and gauge-like symmetries
that \(H_{\sf M}\) possesses and \(H_{\sf AI}\) lacks. To understand
this better, we need to study the effect this mapping has
on relations beyond local anticommutation. Let us
consider first its effect on the extensive relation of Eq. \eqref{ext_r}. We have that
\begin{eqnarray}
Q=\prod_\r {\cal P}_\r&\mapsto& \prod_\r \sigma^x_\r=Q_S,\label{e1}\\
\alpha\prod_\l (ic_{\l1}c_{\l2})&\mapsto& 
\alpha\prod_\l(\eta_\l\sigma^z_\r\sigma^z_{\r'})=\alpha\eta\mathds{1}.
\label{e2}
\end{eqnarray}
As for periodic boundary conditions the left-hand sides of 
Eqs. \eqref{e1} and \eqref{e2} represent the same operator,
but the right-hand sides are different operators,
the mapping as it stands does not preserve the relation of Eq. \eqref{ext_r}.
We know of a solution to this shortcoming from the previous Section.
If we modify one and only one bond placed on some fixed but 
arbitrary link \(\l_0\) to read
\begin{equation}\label{mib}
\alpha\eta\eta_{\l_0}\sigma^z_\r\sigma^z_{\r'}Q_S,
\end{equation}
then
\begin{eqnarray}
&&\alpha\prod_\l (ic_{\l1}c_{\l2})\mapsto  \nonumber \\
&&\alpha^2(\eta\eta_{\l_0}\sigma^z_{\r_0}\sigma^z_{{\r_0}'})Q_S
\prod_{\l\neq\l_0}(\eta_\l\sigma^z_\r\sigma^z_{\r'})=Q_S,
\end{eqnarray}
as required by Eq. \eqref{ext_r}. 

The presence of the modified bond at 
\(\l_0\) introduces a new feature
into the discussion leading to Eq. \eqref{loopc}. Now we have that,
for any elementary loop \(P\),
\begin{equation}\label{modloopc}
\prod_{\l\in P}(\eta_\l\sigma^z_\r\sigma^z_{\r'})=
\left\{
\begin{array}{lcr}
\mathds{1}\prod_{\l\in P}\eta_\l, & \  \ \mbox{if}\ \  & \l_0\not \in P,\\
 \\
\alpha\eta Q_S\prod_{\l\in P}\eta_\l & \ \ \mbox{if}\ \  & \l_0 \in P.
\end{array}\right.
\end{equation}
If we consider the role of the elementary loops  \(P\) in the Majorana system 
\(H_{\sf M}\), and consider the mapping of Eq. \eqref{ispinb}, we see that the 
local symmetries (see Section \ref{sym_section})
\begin{equation}\label{mloopsym}
G_P\equiv \prod_{\l\in P} (ic_{\l1}c_{\l2})
\end{equation}
of \(H_{\sf M}\) are mapped to one of the two possibilities listed in 
Eq. \eqref{modloopc}, showing that as it stands, the mapping of Eq. \eqref{ispinb}
is still not an isomorphism of bond algebras. The problem is that a large number of 
distinct symmetries are being mapped either to a trivial symmetry (a multiple
of the identity operator), or a multiple of the global \(\mathds{Z}_2\) symmetry
\(Q_S\) of the annealed Ising model. We can fix this problem by decomposing
the Hamiltonians  \(H_{\sf M}\) and \(H_{\sf AI}\) into their symmetry sectors, 
 where the obstruction
to the duality mapping disappears. Thus we are able to establish
{\it emergent dualities},\cite{bondprl,ADP} that is, dualities that emerge between
{\it sectors} of the two theories.

The sector decomposition is simple for \(H_{\sf AI}\), that has only one symmetry 
\(Q_S\), with eigenvalues \(q_S=\pm 1\).
Then we can decompose the Hilbert space \({\cal H}_{\sf AI}\)  as
\begin{equation}
{\cal H}_{\sf AI}=\bigoplus_{q_S=\pm 1} {\cal H}_{q_S},
\end{equation}
so that if \(\Lambda_{q_S}\) is the orthogonal projector onto \({\cal H}_{q_S}\), 
then
\begin{equation}
Q_S\Lambda_{q_S}=\pm \Lambda_{q_S}.
\end{equation}

For \(H_{\sf M}\), since its symmetries form a commuting set, one can 
simultaneously diagonalize them and break the Hilbert space \({\cal H}_{\sf M}\) 
into sectors labelled by the symmetries' simultaneous eigenvalues,
\(q=\pm1\) for the global symmetry and \(\Gamma_P=\pm1\) for the loop symmetries:
\begin{equation}
{\cal H}_{\sf M}=\bigoplus_{q,\{\Gamma_P\}} {\cal H}_{q,\{\Gamma_P\}}.
\end{equation}
The Hamiltonian \(H_{\sf M}\) is block-diagonal relative to this decomposition, and,
if \(\Lambda_{q,\{\Gamma_P\}}\) is the orthogonal projector onto the subspace
\({\cal H}_{q, \{\Gamma_P\}}\), we have that 
\begin{eqnarray}
Q\Lambda_{q,\{\Gamma_P\}}&=&q\Lambda_{q,\{\Gamma_P\}},\\
G_P\Lambda_{q,\{\Gamma_P\}}&=&\Gamma_P\Lambda_{q,\{\Gamma_P\}},
\end{eqnarray}
for any elementary loop \(P\).

The problem now is to decide which choice of sectors will make
the projected Hamiltonians  \(H_{\sf M}\Lambda_{q,\{\Gamma_P\}}\) and 
\(H_{\sf AI}\Lambda_{q_S}\) dual to each other. From Eqs. \eqref{e1},
and \eqref{modloopc}, we obtain the relations  
\begin{eqnarray}
q&=&q_S,\label{global}\\
\Gamma_P&=&
\left\{
\begin{array}{lcr}
\prod_{\l\in P}\eta_\l, & \ \ \mbox{if}\ \  & \l_0\not \in P,\\
 \\
\alpha\eta q_s\prod_{\l\in P}\eta_\l & \ \ \mbox{if}\ \  & \l_0 \in P,
\end{array}\right.\label{local}
\end{eqnarray}
which allow us to connect the two theories
\begin{equation}\label{MQW}
U_\d H_{\sf M}\Lambda_{q,\{\Gamma_P\}} U_\d^\dagger =
H_{\sf AI}\{\eta_{\l}\} \Lambda_{q_S},
\end{equation}
where the unitary transformation \(U_\d\) implements an 
{\it emergent duality} that holds only on the indicated sectors of the
two theories. 

The dual spin representation 
of  \(H_{\sf M}\) projected onto the gauge-invariant 
sector \(q=1,\{\Gamma_P=1\}\) is given by 
the inhomogeneous Ising model (\(\eta_\l=1\) on every link)
\begin{equation}
H_{\sf AI}\{1\}=-\sum_{\l} J_{\l} \sigma^z_\r\sigma^z_{\r'}-\sum_{\r} h_\r \sigma^x_\r,
\end{equation}
and is known as a {\it gauge-reducing duality}. \cite{ADP} 
For the special case of the square lattice and homogeneous couplings, 
one would expect that this sector contains the ground state of $H_{\sf M}$. 
This latter result was derived, using methods very different to ours, 
in Ref. \onlinecite{Terhal}. 

\subsection{Physical Consequences}
\label{conseq}

We have by now seen, on general networks in an arbitrary number of dimensions, 
that ordinary quantum Ising gauge theories (and their generalizations) and 
annealed transverse-field Ising models arise from the very same Majorana system 
when it is dualized in different ways. Therefore, by transitivity,
\begin{eqnarray}
H_{\sf QIG} \stackrel{\sf dual}{\longleftrightarrow} H_{\sf AI}.
\label{QVAI}
\end{eqnarray}
This correspondence leads to several consequences. In its simplest incarnation, that 
for $D=2$ Majorana networks, this duality connects, via an imaginary-time 
transfer matrix (or $\tau$-continuum limit) approach, \cite{kogut,ADP}
disordered $D=3$ classical Ising models to $D=3$
classical Ising gauge theories. 
In its truly most elementary rendition among these planar networks,
that of the square lattice, the duality of Eq. (\ref{QVAI}) implies that the effect
of the bimodal annealed disordering fields $\eta_{\l} = \pm 1$ is immaterial 
in determining the universality class of the system. This is so as the standard 
random transverse-field Ising model on the square lattice 
\begin{eqnarray}
\label{tif}
H_{\sf RTFIM}=-\sum_{\l} J_{\l}\sigma^{z}_{\r} 
\sigma^{z}_{\r'}-\sum_{\r} h_{\r}\sigma^{x}_{\r}
\end{eqnarray}
(i.e., Eq. \eqref{hspin} in the absence of annealed bimodal disorder)
similarly maps, via a transfer matrix approach, onto a corresponding classical 
Ising model on a cubic lattice. The uniform transverse-field Ising model 
(that with uniform $J_{\l}$ and $h_{\r}$) maps onto the uniform $D=3$ Ising model. 
Thus, in this latter case, the {\it extremely disordered} system with annealed 
random exchange constants exhibits the standard $D=3$ Ising type behavior of uniform 
systems.

By the dualities of Sections \ref{tdn} and \ref{ttdn},  general  
multi-particle, or multi-spin, spatio-temporal correlation functions in 
different systems can be related to one another. In particular, by  
Eq. \eqref{ppcc} relating the Majorana system with the quantum Ising gauge
theory, the two correlators 
\begin{eqnarray}
\langle \prod_{\r,\l} {\cal{P}}_{\r} (t) (ic_{\l1}c_{\l2})(t') \rangle =
\langle \prod_{\r,\l}\widetilde{\cal{P}}_\r(t) \sigma^{x}_{\l}(t') \rangle
\end{eqnarray}
are equal. Thus, if certain correlators 
(e.g., standard static two-point correlation functions,
autocorrelation functions, or four-point correlators such as those prevalent in the 
study of glassy systems) \cite{4-point}
appear in the spin systems, then dual correlators appear in the interacting 
Majorana system with identical behavior. 
An exact duality preserves the equations of motion, and so the
dynamics of dual operators are the same. \cite{ADP} 
Similarly, by the duality of Eq. \eqref{MQIG}, the phase diagrams describing the 
Majorana networks are identical to those of quantum Ising gauge systems. 
In instances in which the quantum Ising gauge theories have been investigated, 
the phase boundaries in the Majorana system may thus be mapped out without further ado. 

Lattice gauge theories with homogeneous couplings, i.e., uniform lattices,  
have been investigated extensively. \cite{wegner, fradkin_susskind}
As we alluded to above, it is well appreciated that the quantum Ising 
gauge theory on a square lattice can be related,
via a Feynman mapping, to an Ising gauge theory on the cubic lattice with the 
classical action 
\begin{equation}
{\cal{S}}_{\sf IG}= -  K \sum_{P} {\cal{P}}_{P}.
\end{equation}
The latter has a transition \cite{IZ} at $K=K_{c} = 
0.761423$, a value dual \cite{wegner} to the 
critical coupling (or inverse critical temperature when the exchange constant is 
set to unity) of the $D=3$ classical Ising model with nearest 
neighbor coupling, $\tilde{K}_{c} = 0.2216595$.
Similar transitions between a confined (small $K$) to a deconfined (large $K$) 
phases appear in general uniform coupling lattice gauge theories with other geometries.
Phase transitions mark singularities of
the free energy, that are always identical in any two dual models. \cite{ADP} 
In our case of interest here, by the correspondence of Eq. (\ref{MQIG}), 
identical transition points must thus appear in the dual Majorana theories. 
In particular, the transition points in the Majorana 
system are immediately determined by their dual spin counterpart. 
More precisely, the Majorana uniform network depicted in Fig. \ref{l12} 
displays a quantum critical point of the $D=3$ Ising universality 
class at $(J/h)_c=-2 \tilde{K}_{c}/\ln \tanh \tilde{K}_{c}=0.29112$.

In theories with sufficient 
disorder (e.g., quenched exchange couplings, fields, or spatially varying 
coordination number), rich behavior such as that exemplified by 
spin glass transitions or Griffiths singularities \cite{Grf} may appear. 
According to Eq. \eqref{MQIG}, in architectures with non-equidistant 
superconducting grains of random 
sizes, the effective couplings $\{J_{\l}\}$ and $\{h_{\r} \}$ are not uniform and 
may lead to spin glass, Griffiths, or other behavior whenever the corresponding 
dual gauge theory exhibits these as well.  
We note that the random transverse field Ising model 
of Eq. \eqref{tif} is well known to exhibit a (quantum) spin glass 
behavior. \cite{dsfisher,schechter} If and when it occurs, glassy 
(or spin-glass) dynamics in the annealed or gauge spin systems will, by our 
mapping, imply corresponding glassy (or spin-glass)  dynamics in the Majorana 
system as well as interacting 
electronic systems (leading to {\it electron glass} behavior). The disordered 
quantum Ising model was employed in the study of the insulator to superconducting
phase transition in granular superconductors. \cite{mezard} 
Numerous electronic systems are indeed non-uniform \cite{electron_nonuniform} 
and/or disordered. \cite{electron_glass} 

\section{Spin Duals to Square Lattice Majorana Systems}

Thus far, we provided a systematic analysis of symmetries and dualities 
for Majorana systems supported on networks in any number of spatial dimensions. 
It is instructive to consider particularly simple architectures as these highlight 
salient features and, on their own merit,
provide new connections among well studied theories. In what follows, we will 
focus on the square lattice superconducting grain array of Fig. \ref{l12},
and some honeycomb and checkerboard lattice spin dual models. 

\subsection{The XXZ Honeycomb Compass Model}

\begin{figure}[h]
\includegraphics[angle=90, width=.45\columnwidth]{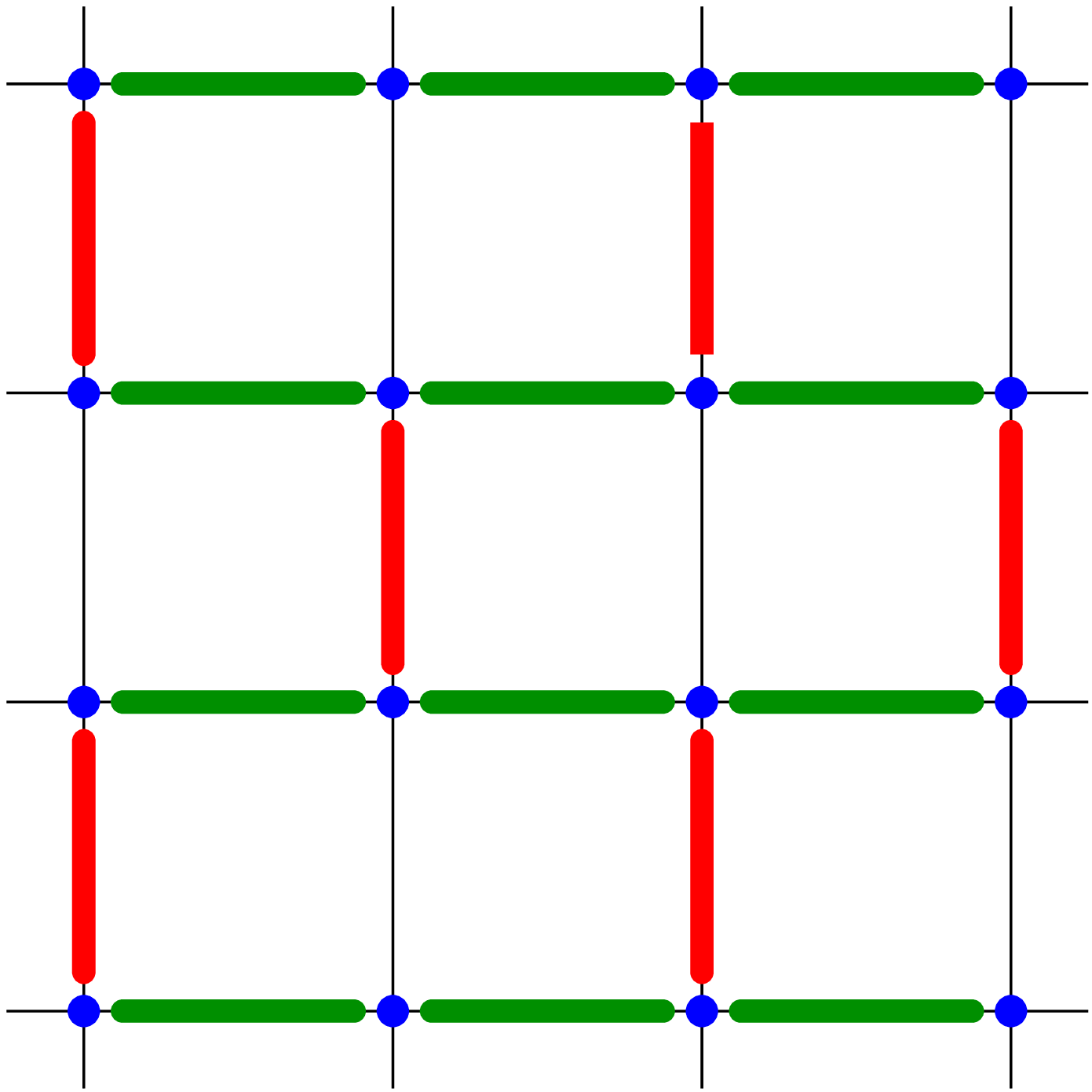}\ \ \ \ \ 
\includegraphics[angle=0, width=.48\columnwidth]{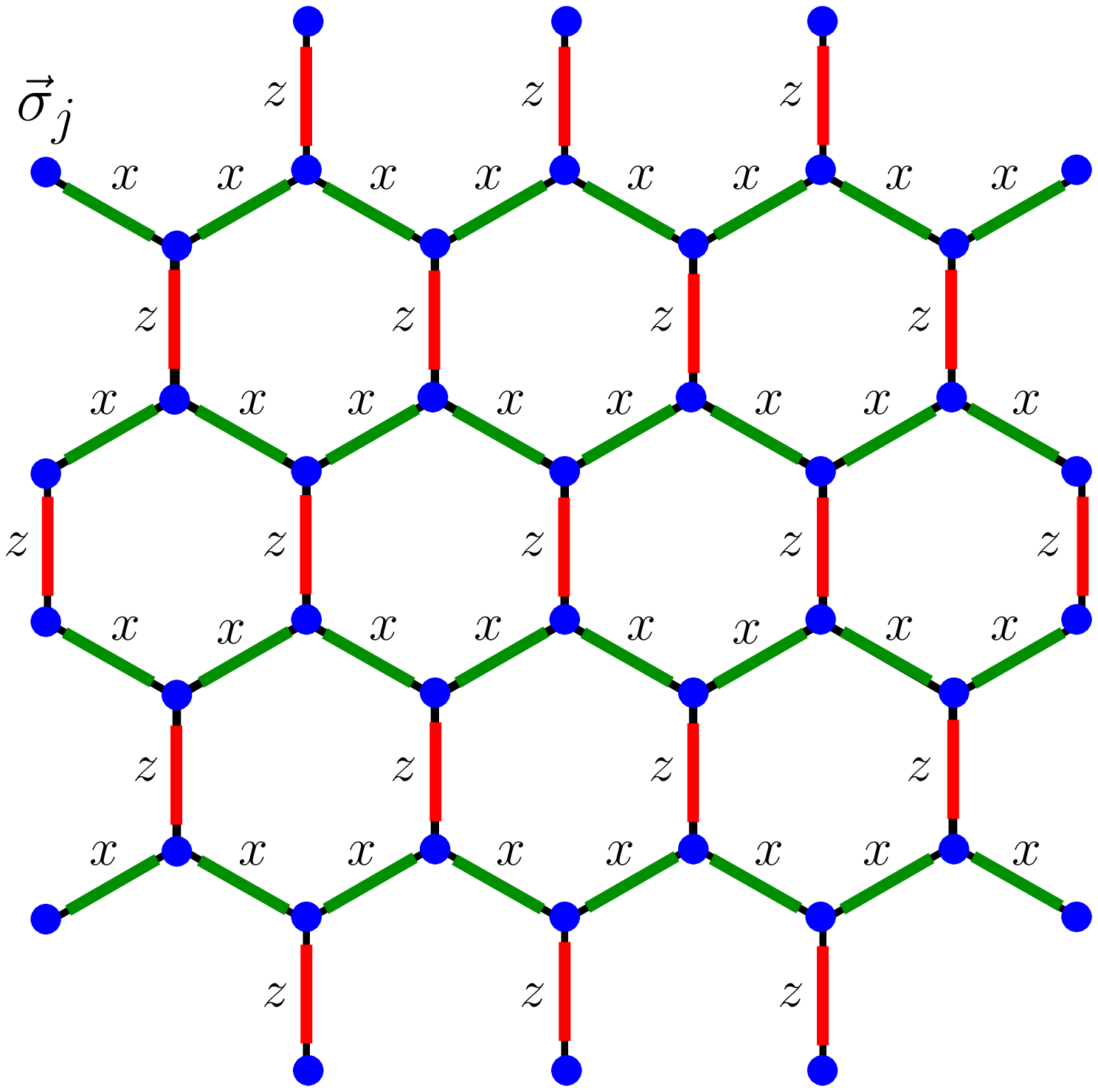}
\caption{The brick-wall planar orbital compass model \cite{ADP} (shown on the left)
can be seen as a simpler relative of the XXZ honeycomb compass model, by placing it 
on a honeycomb
lattice as shown on the right.}
\label{boc}
\end{figure}

The Majorana system \(H_{\sf M}\) of Eq. (\ref{Ham}) in a square lattice 
is dual to a very interesting
spin Hamiltonian on the honeycomb lattice, see Fig. \ref{boc}. 
The dual spin model may be viewed as
an intermediate between the classical Ising model on the honeycomb lattice 
(involving products of a single spin component $(\sigma^{z}$) between 
nearest neighbors) and Kitaev's honeycomb model, \cite{Kitaev06} for which
the bonds along the three different directions in the lattice are
respectively pairwise products
of the three different spin components. This particular spin Hamiltonian,
which we dub {\it XXZ honeycomb compass model}, is described by
\begin{eqnarray}
\label{xxz}
H_{\sf XXZh} = - \!\!\!\!\! \sum_{\mbox{\sf \tiny non-vertical~links}} \!\! J_{\l} \, \sigma^{x}_{\r} 
\sigma^{x}_{\r + \hat{e}_{\l}} 
-\!\!\! \sum_{\mbox{\sf \tiny vertical~links}} \!\! h_{\r} \, \sigma^{z}_{\r} \sigma^{z}_{\r + \hat{e}_{z}} ,
\end{eqnarray}
where each $S=1/2$ is located on the vertices $\r$ of a honeycomb lattice, 
 and $\sigma^{x,z}_{\r}$ are the corresponding
Pauli matrices. The qualifier  ``non-vertical links'' alludes to the two 
diagonally oriented directions of the honeycomb lattice
while ``vertical links'' are, as their name suggests, the links parallel to the 
vertical direction in Fig. \ref{boc}. The unit vector $\hat{e}_{\l}$ points along 
the diagonal link $\l$ and may be oriented along any of the two diagonal directions.
The XXZ honeycomb compass model exhibits local symmetries associated with every lattice 
site $\r$, 
\begin{eqnarray}
G_{\r}^{\sf XXZh} =  \sigma^{x}_{\r} \sigma^{x} _{\r + \hat{e}_{z}}.
\end{eqnarray}
Similarly, the XXZ system exhibits $d=1$ symmetries of the form
\begin{eqnarray}
Q^{\sf XXZh}_{\ell} = \prod_{\r \in \ell} \sigma^{z}_{\r}
\end{eqnarray}
associated with every non-vertical contour $\ell$ (i.e., that composed of the diagonal non-vertical links) 
that circumscribes one of the toric cycles.
  
We provide, in the left-hand panel of Fig. \ref{boc}, a simple schematic
of the topology of the honeycomb lattice - that of a ``brick-wall lattice''. \cite{orbital,CN} 
The brick-wall lattice also captures the connections in the honeycomb lattice.
It is formed by the union of the highlighted vertical (red) and 
horizontal (green) links in the left-hand side Fig. \ref{boc}. 
The brick-wall lattice can be obtained by ``squashing''  the honeycomb lattice 
to flatten its diagonal links while leaving
its topology unchanged in the process.  In the brick-wall lattice, $\hat{e}_{\l}$ 
simply becomes a unit vector along 
the horizontal direction. 
As can be seen by examining either of the panels of Fig. \ref{boc},  
the centers of the vertical links of the honeycomb (or brick-wall) lattice form, up 
to innocuous dilation factors, a square lattice. As is further evident on 
inspecting Fig. \ref{boc}, between any pair of 
centers of neighboring vertical (red) links, there
lies a center of a non-diagonal (green) link. 
This topological connection underlies the duality between the Majorana model
on the square lattice and the XXZ honeycomb compass spin model.
We explicitly classify the bonds in the 
Hamiltonian of Eq. (\ref{xxz}) related to the
two types of geometric objects: 
\begin{enumerate}
\item 
Bonds of type ({\sf i}) are associated with the products $\{\sigma^{x}_{\r} 
\sigma^{x}_{\r + \hat{e}_{\l}}\}$ on diagonal links of the lattice. 
They each anticommute with two
\item
Bonds of type ({\sf ii}), affiliated with products $\{\sigma^{z}_{\r} 
\sigma^{z}_{\r+\hat{e}_{z}}\}$ on the vertical links. Each one of these 
bonds anticommute with four bonds of  type ({\sf i}).
\end{enumerate}

We merely note 
that replacing the bonds of the Majorana model on a square lattice, 
as they appear in 
the bond algebraic relations (1-3) 
of Section \ref{main}, by the ones above 
leads to three equivalent relations  that 
completely specify the bond algebra of
the system of Eq. (\ref{xxz}). 
As we have earlier seen also the quantum Ising gauge theory of Eq. (\ref{QIG})
and the annealed transverse-field Ising model of Eq. (\ref{hspin}) have bonds that 
share the same three basic bond algebraic relations.  
Thus we conclude that the
XXZ honeycomb compass model is exactly dual to the quantum Ising gauge theory 
of Eq. (\ref{QIG}) on the square lattice. In its uniform rendition 
(with all couplings $J_{\l}$ and fields $h_{\r}$ being spatially
uniform) {\it the XXZ honeycomb compass system lies in the 3D Ising
universality class}. Similarly, many other properties 
of the XXZ honeycomb compass model can be inferred from the heavily investigated quantum 
Ising gauge theory. 

The duality between the XXZ honeycomb compass model 
and its Majorana system equal on the square
lattice affords an example of a duality in which the Hilbert space size is preserved 
as we now elaborate. The XXZ  theory of Eq. (\ref{xxz}) is defined
on a Hilbert space of size $\dim {\cal{H}}_{\sf XXZh} = 2^{N_{\sf hl}}$ where 
$N_{\sf hl}$ is the number of sites on the honeycomb lattice while that of 
the Majorana model of Eq. (\ref{Ham})
was on a Hilbert space of dimension $\dim{\cal{H}}_{\sf M} = 4^{N_{\r}}$. Now, 
for a given number $N_{\r}$ of vertical links on the honeycomb lattice, we have
the same number of bonds of type ({\sf i}) and ({\sf ii}) as we had in the Majorana 
system  while having $N_{\sf hl} = 2N_{\r}$ lattice sites. 

\subsection{Checkerboard  model of $(p+ip)$ superconducting grains}
\label{xmcd}

In Ref. \onlinecite{XM},  Xu and Moore, motivated by an earlier work of Moore 
and Lee, \cite{ML} proposed the following spin Hamiltonian 
 \begin{equation}
 \label{xme}
H_{\sf XM}= - \sum_\r (h^{\sf XM}_{\r} \sigma^x_\r + J_{\Box}^{\sf XM} \square\sigma^z_\r)
\end{equation}
to describe the time-reversal symmetry breaking characteristics in a matrix 
of unconventional $p$-wave granular superconductors on a square lattice. 
In writing Eq. \eqref{xme}, we employ a shorthand 
\begin{equation}
\label{xml}
\square\sigma^z_\r \equiv \sigma^z_\r\sigma^z_{\r+\i}\sigma^z_{\r+\i+\j}\sigma^z_{\r+\j},
\end{equation}
to denote the square lattice plaquette product,
where $\i$ and $\j$ denote unit vectors along the principal lattice directions. 
It is important to emphasize that
the spins ${\bf{\sigma}}_{\r}^{x,z}$ in Eqs. \eqref{xme}, \eqref{xml} 
are situated at the vertices $\r$ of the square lattice
(not on the links (or link centers) as in gauge theories). 
The eigenvalues 
$\sigma^{z}_{\r}=\pm 1$ describe whether the superconducting
grain located at the vertex of the square lattice $\r$ has a $(p + ip)$ or a $(p - ip)$ order
parameter. 
\begin{figure}[h]
\centering
\includegraphics[angle=0, width=.8\columnwidth]{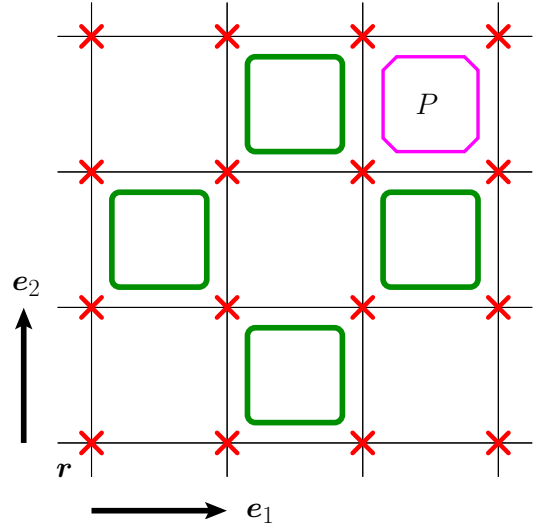}\ \ \  
\caption{The checkerboard Xu-Moore (CXM) model of Eq. (\ref{checkerboard}). 
The symmetry plaquettes $P$ constitute half 
of all the plaquettes of the lattice, while the interaction plaquettes 
$\square\sigma^z_\r$ represent the other half.}
\label{sxm}
\end{figure}

We show next that a $D=2$ {\it checkerboard} rendition of the XM model which we denote by CXM
(see Fig. \ref{sxm}) is dual to the Majorana system on the square lattice (which is, as we showed, 
dual to the XXZ honeycomb compass model and all of the other models that we discussed
earlier in this work). This system is defined by the following Hamiltonian 
\begin{equation}\label{checkerboard}
H_{\sf CXM}= - \sum_\r h_{\r} \sigma^x_\r- \sum_{x_{1}+x_{2}={\sf odd}} 
J_{\Box}^{\sf XM} \square\sigma^z_\r.
\end{equation}
In this system, the plaquette operators $\square\sigma^z_\r$ (with \(\r=x_{1}\i+x_{2}\j\))
appear in every other plaquette (hence the name ``checkerboard''). These 
plaquettes are present only if  \(x_{1}+x_{2}\) is an odd integer as emphasized in 
Eq. (\ref{checkerboard}). The model has the following local symmetries
\begin{eqnarray}
G_P=\prod_{\r \in P} \sigma^x_\r ,
\end{eqnarray}
where $P$ are those plaquettes appearing whenever \(x_{1}+x_{2}\) is an even integer.

The proof of our assertion above concerning the duality of this system to the Majorana system 
of Eq. (\ref{Ham}) when implemented on the square lattice is straightforward and will
mirror, once again, all of our earlier steps. We may view the Hamiltonian of 
Eq. (\ref{checkerboard}) as comprised of two basic types of bonds: \newline
\begin{enumerate}
\item
Bonds of type ({\sf i}) are on-site operators $\{\sigma^{x}_{\r}\}$  associated  
with local transverse fields. 
\item
Bonds of type  ({\sf ii}) are the plaquette product operators 
$\{\square\sigma^z_\r\}$ of Eq. \eqref{xml}, 
for plaquettes whose bottom left-hand corner $\r$ is an ``odd'' site.
\end{enumerate}

\begin{figure}[h]
\centering
\includegraphics[angle=0, width=.47\columnwidth]{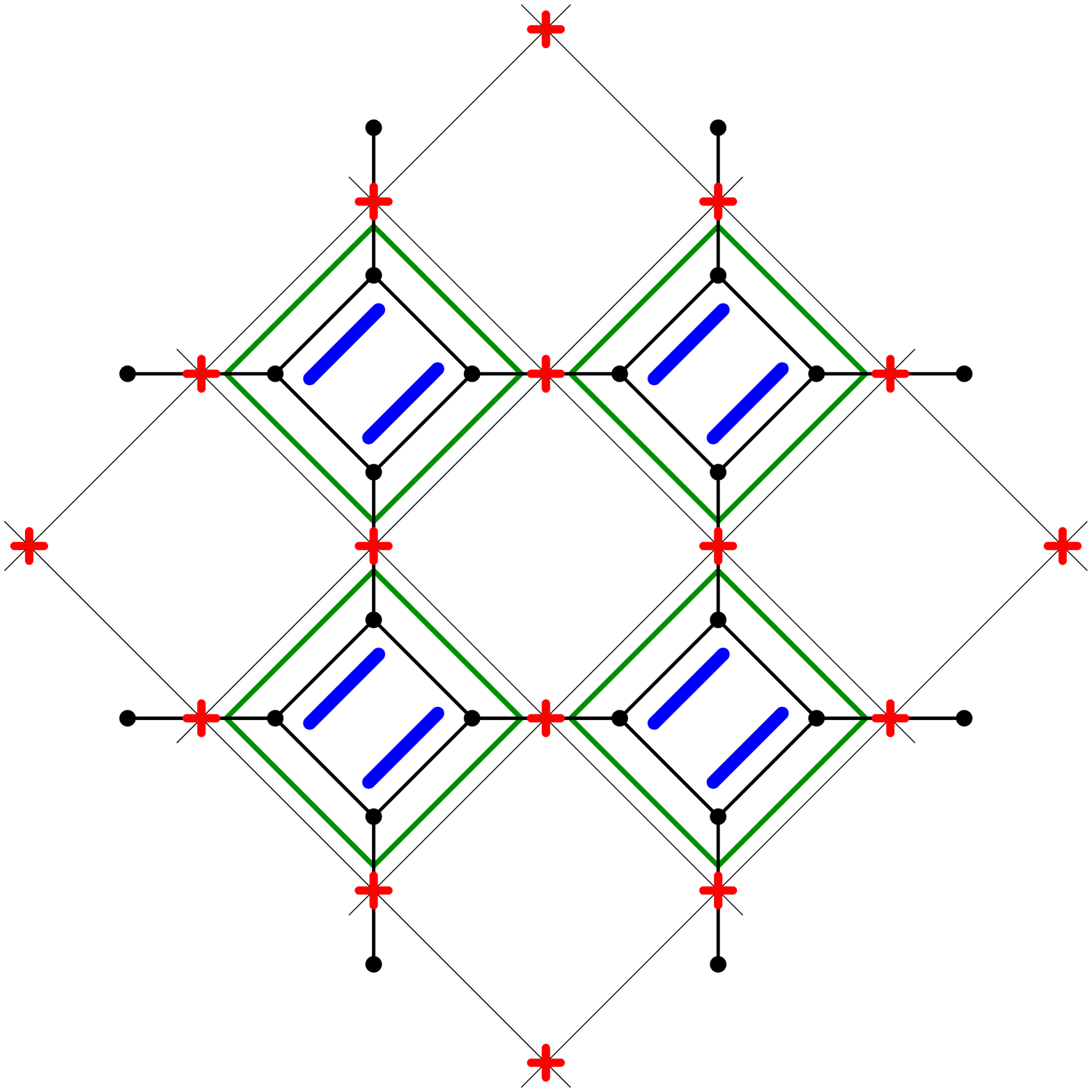}\ \ \  
\includegraphics[angle=0, width=.47\columnwidth]{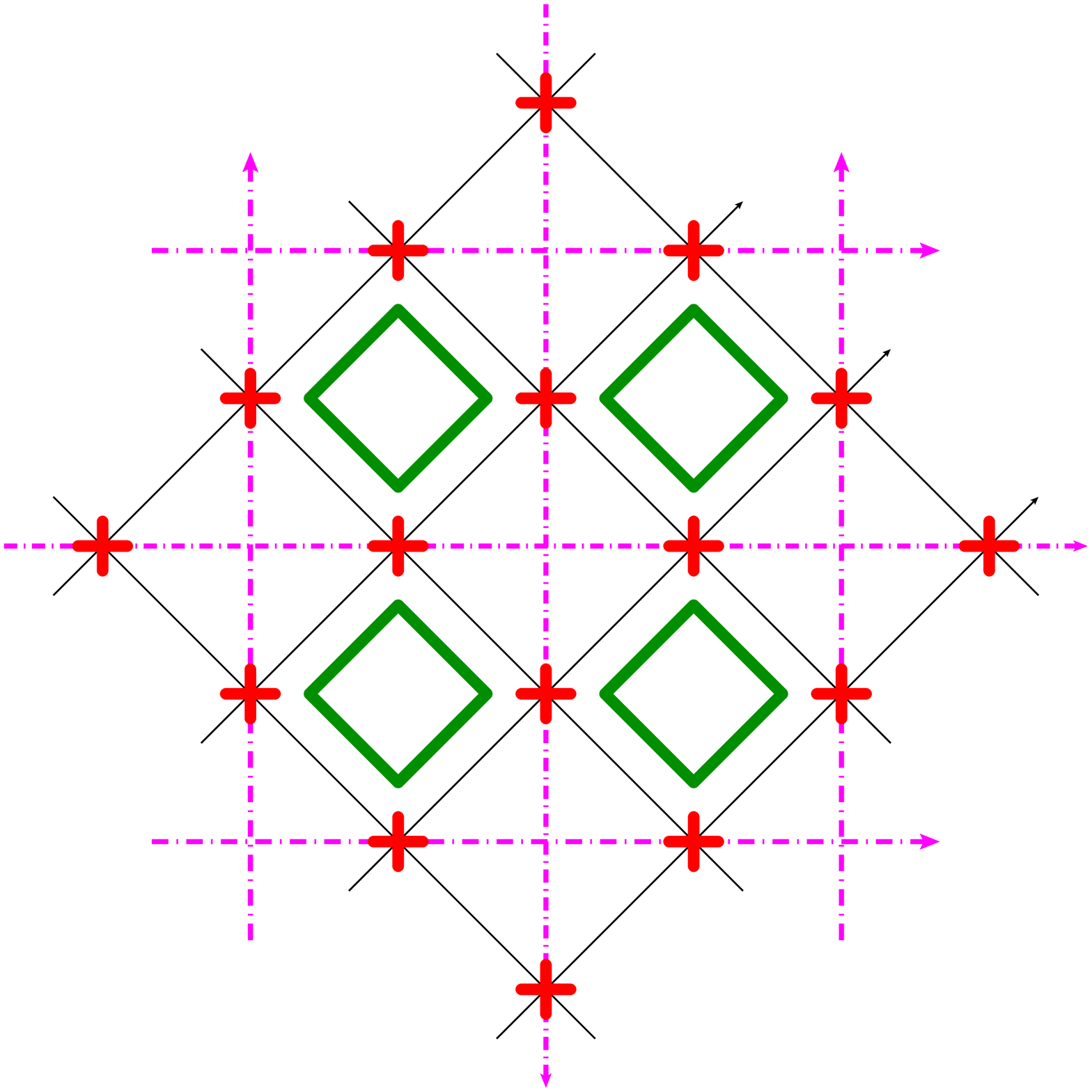}
\caption{The $D=2$ checkerboard Xu-Moore (CXM) model is dual to the 
Majorana system in a square lattice as shown on the left. On
the right, we rotate and redefine the lattice in a manner which highlights its connection 
to the quantum Ising gauge (QIG) theory of Eq. (\ref{QIG}). }
\label{sxmnew}
\end{figure}

The basic network structure underlying these bonds is simple and, apart from 
an interchange of names, identical to that of the Majorana system on the square 
lattice of Fig. \ref{l12}
as well as that of the XXZ honeycomb compass model of Fig. \ref{boc}. To see this, we note 
that in the checkerboard of Fig. \ref{sxm}, the four-fold coordinated interaction
plaquettes generate, on their own,
a square lattice grid. Between any two neighboring interaction plaquettes on this 
square lattice array, there is a lattice site $\r$ (see Fig. \ref{sxmnew}). As in our earlier proof of 
the duality, 
we simply remark that replacing the bonds of the Majorana model on a square lattice, 
as they appear in 
the bond algebraic relations (1-3) 
of Section \ref{main}, by the ones above 
leads to three equivalent relations  that 
completely specify the bond algebra of
the CXM system. The Majorana and CXM models 
are thus dual to one another 
($H_{\sf M} \leftrightarrow H_{\sf CXM}$)
when their couplings are related via the correspondence
\begin{eqnarray}
J_{\l} \leftrightarrow h^{\sf XM}_{\r}, \nonumber
\\   h_{\r} \leftrightarrow J_{\Box}^{\sf XM}.
\end{eqnarray}
Thus, the CXM model joins
the fellowship of all other dual theories (with the same network 
connectivity) that we discussed in this work (i.e., the Majorana, quantum Ising 
gauge, and annealed transverse field Ising models
on the square lattice as well as the XXZ compass model on the honeycomb (or equivalent 
brick-wall) lattice). 

On the right-hand half of Fig. \ref{sxmnew}, we pictorially illustrate 
the connection between the CXM model and the quantum Ising gauge theory. The 
individual sites of the checkerboard lattice of Fig. \ref{sxm}
(the sites at which the local transverse fields 
are present) map onto links of the gauge theory (Section \ref{tdn}).
Similarly, the interaction plaquettes of the CXM model  map 
into plaquettes of the quantum Ising gauge theory. Note, on the 
right, that as is geometrically well appreciated, 
the four center-points of the individual links 
on the square (gauge theory) lattice can either circumscribe 
interaction plaquettes of the gauge theory  or may correspond to four links 
that share a common endpoint that do  form a ``star'' configuration. \cite{ADP}
 In 
particular, by its duality to the quantum Ising theory, the CXM 
rigorously lies in the 3D Ising universality class
when the couplings $J_{\Box}^{\sf XM}$ and $h^{\sf XM}_{\r}$ are spatially uniform. 
For a given equal number of bonds in both the Majorana system and the CXM theory, it is
readily seen that the Hilbert space dimensions of both theories are the same, 
$\dim {\cal{H}}_{\sf M} = \dim{\cal{H}}_{\sf CXM}$.

\section{
Simulating Hubbard-like models with Majorana networks}

The Dirac, fermionic, annihilation and creation operators,
$\{d_\r\}$ and $\{d_\r^{\dagger}\}$ respectively, can be expressed as a 
linear combination of two Majorana fermion operators. For example, 
if we are interested in {\it two-flavor} Dirac operators a possible realization is 
(see Fig. \ref{l12})
\begin{eqnarray}
\label{mapfmm2}
d_{ \r \uparrow} &=& \frac{1}{\sqrt{2}}(c_{\l_{1}1} + i c_{\l_{3}2}), \ 
d^{\dagger}_{\r \uparrow} = \frac{1}{\sqrt{2}}(c_{\l_{1}1} - i c_{\l_{3}2}), \nonumber
\\ 
d_{ \r \downarrow} &=& \frac{1}{\sqrt{2}}(c_{\l_{2}1} + i c_{\l_{4}2}), \ 
d^{\dagger}_{\r \downarrow} = \frac{1}{\sqrt{2}}(c_{\l_{2}1} - i c_{\l_{4}2}),
\end{eqnarray}
where $\r \in \l_1,\l_2,\l_3,\l_4$.

A system of interacting Dirac fermions (e.g., electrons) on a general graph 
can be mapped onto that of twice the number of Majorana fermions on the same graph,
and each Dirac fermion
is to be replaced by two Majorana fermions following the substitution of
Eq. (\ref{mapfmm2}). Thus, any granular system of the form
of Eq. (\ref{Ham}) in which each grain $\r$ has $q_{\r}= 2z_{\r}$ 
neighbors, can be mapped onto a Dirac fermionic system on the same
graph in which on each grain there are $z_{\r}$ Dirac fermions. 
There are many possible ways to pair up the Majorana fermions in the system
of Eq. (\ref{Ham}) 
to yield a corresponding system of Dirac fermions. Equation  (\ref{mapfmm2}) represents
just one possibility. Another possible way to generate (spinless) Dirac 
fermions is
\begin{eqnarray}
\label{mapfmm}
d_\l = \frac{1}{\sqrt{2}}(c_{\l1} + i c_{\l2}), \ 
d^{\dagger}_{\l} = \frac{1}{\sqrt{2}}(c_{\l1}-i c_{\l2}).
\end{eqnarray}
{\it All of the spin duals that we derived for Majorana 
fermion systems hold, {\it mutatis mutandis}, for these
systems of Dirac fermions on arbitrary graphs.} 
In this sense, dualities afford an
alternative, flexible approach to fermionization that does not rely on the Jordan-Wigner
transformation. \cite{ADP} Most importantly, one can use these mappings to 
simulate models of strongly interacting Dirac fermions, such as Hubbard-like models, 
on the experimentally realized Majorana networks. In other words, one can 
engineer {\it quantum simulators} out of these Josephson junction arrays. 

As a concrete example, we consider the square-lattice array of Fig. \ref{l12}
and transform, on this lattice,
the Majorana system of Eq. (\ref{Ham}) into a {\it two-flavor Hubbard model with 
compass-type pairing and hopping}.
Based on our analysis thus far  we will
illustrate that this variant of the 2D Hubbard model is exactly dual to the 2D 
quantum Ising gauge theory and thus 
lies in the 3D Ising universality class. 
Consider the mapping of Eq. \eqref{mapfmm2}. 
With $n_{\r \sigma} = d_{\r \sigma}^{\dagger} d_{\r \sigma}$ 
($\sigma=\uparrow, \downarrow$), a Hubbard type term with 
on-site repulsion $U_{\r}$ becomes
\begin{eqnarray}
\label{urh}
U_{\r} (n_{\r \uparrow}-1) (n_{\r \downarrow}-1)= U_{\r} ({\cal P}_\r -1) ,
\end{eqnarray}
akin to the second term of Eq. (\ref{Ham}) with $h_{\r} \leftrightarrow U_{\r}$ 
(up to an irrelevant constant). 
In what follows we assume that the network array of Fig. \ref{l12} has unit 
lattice constant.

The Majorana bilinear that couples, for instance, 
the bottom most corner of the grain that 
is directly above $\r$
(i.e., site ${\r + \j}$)
to the top-most site of grain $\r$ (with thus a link $\l$ that is vertical) becomes
\begin{eqnarray}
-i J_{\l} c_{\l_{2}1} c_{\l_{2}2} 
= \frac{J_{\l}}{2}
(d_{\r \downarrow}^{\dagger}+d_{\r \downarrow} )
( d_{\r + \j \downarrow}^{\dagger}-d_{\r + \j \downarrow}).
\label{jccj}
\end{eqnarray}
Similarly, for horizontal links $\l$, the bilinear in the first term of 
Eq. \eqref{Ham} realizes pairing hopping terms involving only the 
$\uparrow$ flavor of the fermions. 
Thus, the Hamiltonian of Eq. (\ref{Ham}) becomes a Hubbard type Hamiltonian with 
bilinear terms containing hopping and pairing terms between electrons of the up or down flavor
for links $\l$ that are vertical or horizontal, respectively. Such a dependence of 
the interactions
between the internal spin flavor on the relative orientation of the two interacting
electrons in real-space bears a resemblance to ``compass type'' systems. \cite{brink}
Putting all our results together, the Dirac fermion Hamiltonian on the square lattice with
pair terms of the form of Eq. (\ref{jccj}) augmented by the on-site Hubbard type interaction
term of Eq. (\ref{urh}) is dual to all of the other models that we considered thus far 
in this work. In particular, as such this interacting Dirac fermion (or electronic) system
{\it is not of the canonical non-interacting Fermi liquid form}. Rather, this 
system lies in the 3D Ising universality class. 

\begin{figure}[th]
\centering
\includegraphics[angle=0,width=.9\columnwidth]{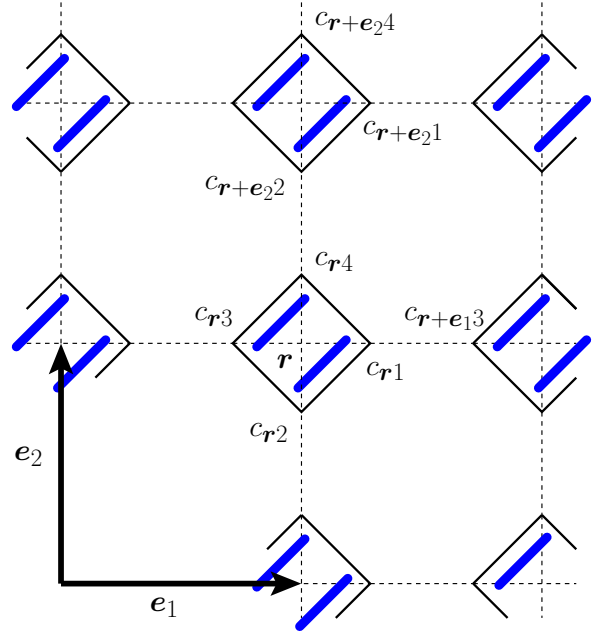}
\caption{A labeling of the Majorana wire endpoints on the square lattice
which we use here to explicitly represent the standard electronic Hubbard 
model in terms of Majorana operators. This is a different 
 labeling than the one in Fig. \ref{l12}. }
\label{majorana_U}
\end{figure}

The standard Hubbard model with $SU(2)$ spin symmetry, which up to 
chemical potential terms is given by ($\alpha=1,2$)
\begin{eqnarray}
H_{\sf Hub} = -t \sum_{\r, \alpha, \sigma} 
(d_{\r \sigma }^{\dagger} d_{\r+{\bm e}_{\alpha} \sigma}  + {\sf h.c.}) \nonumber
\\ + U \sum_{\r} (n_{\r \uparrow}-1) (n_{\r \downarrow}-1) ,
\end{eqnarray}
can be written as a sum of terms of the form of 
Eq. \eqref{urh}
augmenting many Majorana Fermi bilinear coupling sites on nearest neighbor grains 
(i.e., $\r$ and $\r \pm {\bm e}_\alpha$).  As we illustrate in Fig. \ref{majorana_U}, we label the four Majorana 
modes on each grain $\r$ as $\{c_{\r a}\}_{a=1}^{4}$.  In terms of these,
the Hubbard Hamiltonian becomes
\begin{eqnarray}
H_{\sf Hub} = &-&t  \sum_{\r, \alpha, a=1,2} i (c_{\r a} 
c_{\r+ {\bm e}_{\alpha} a+2} +  
c_{\r+ {\bm e}_{\alpha} a}c_{\r a+2}) \nonumber \\
&+& U \sum_{\r} ({\cal P}_\r -1).
\end{eqnarray}

Thus, the Hubbard Hamiltonian may be simulated via Majorana wires with multiple Josephson junctions. 

Appendix \ref{appA} describes the possible simulation of quantum spin 
$S=1/2$ systems in terms of Majorana networks. 

\section{Conclusions} 
We conclude with a brief synopsis of our findings. This work focused on the 
interacting Majorana systems of Eq. (\ref{Ham}) {\it on general lattices and 
networks}. By employing the standard representation of Dirac fermions as a 
linear combination of Majorana fermions, our results similarly hold for a general 
class of interacting Dirac 
fermion systems 
on general graphs. Towards this end, we heavily invoked two principal tools:

\begin{itemize}
\item
The use of $d$-dimensional gauge-like symmetries that mandate 
dimensional reduction and 
TQO via correlation function bounds. \cite{TQO,BN,ads}
These symmetries lead to bounds on the autocorrelation times. \cite{ads}

\item
The bond-algebraic theory of dualities 
\cite{ads,NO,bond,orbital,bondprl,ADP,clock} as it, 
in particular, pertains to very general dualities and fermionization 
\cite{bondprl,ADP} to 
obtain multiple exact spin duals to these systems, in arbitrary dimensions 
and boundary conditions,  
and for finite or infinite systems. 
\end{itemize}

Using this approach, we demonstrated that 
\begin{itemize}
\item
The Majorana systems of Eq. (\ref{Ham}), 
standard quantum Ising gauge theories (Eq. (\ref{QIG})) 
and,  transverse-field
Ising models with annealed bimodal disorder (Eq. (\ref{hspin})) 
are all dual to one another on general lattices and networks. 
The duality afforded
an interesting connection between heavily disordered annealed Ising systems and uniform 
Ising theories. The spin duals further enable us
to suggest and predict various transitions as well as spin-glass type behavior in 
general interacting Majorana fermion (and Dirac fermion) systems. 
The representation of Dirac fermions via Majorana fermions enlarges the scope of our results.
In particular, as Eq. (\ref{urh}) makes evident, the standard on-site Hubbard term in electronic systems
is exactly of the same form as that of the intra-grain coupling in the interacting Majorana systems
that we investigated. We similarly represented the bilinear in the Majorana model of Eq. (\ref{Ham}) as a 
Dirac fermion form (Eq. (\ref{jccj})). Following our dualities, on the square lattice, 
the interacting Dirac fermion (or electronic) Hamiltonian formed by the sum of all terms of the  
form of Eqs. (\ref{urh}, \ref{jccj}) is dual to the quantum Ising gauge theory 
and thus lies in the 3D Ising universality class, notably different from standard 
non-interacting Fermi liquids; this non-trivial electronic system features 
Hubbard on-site repulsion augmented by ``compass'' type hopping and pairing terms.  We further 
showed how to simulate bona fide Hubbard type electronic Hamiltonians via Majorana wire networks. 

\item
Several new systems were introduced and investigated via the use of bond 
algebras: \newline
(1) the ``XXZ honeycomb compass'' model of 
Eq.  (\ref{xxz}) (a model intermediate between the classical Ising model on the 
honeycomb lattice
and Kitaev's honeycomb model and,  \newline
(2) a checkerboard version of the Xu-Moore model 
for superconducting $(p+ip)$ arrays (Eq. (\ref{checkerboard})). \newline
By the use 
of dualities, we illustrated that
both of these systems lie in the 3D Ising universality class.  
\end{itemize}

As evident in our work, all of the considerations necessary to attain these results 
were, to say the least, very simple by comparison to other approaches to duality 
that generally require far more involved calculations.
In the appendices we discuss 
other connections between Majorana and spin systems.

\section{Acknowledgments}

This work was partially supported by NSF CMMT 1106293
at Washington University.   \newline

\appendix

\section{Dualities in finite systems with open boundary conditions}
\label{appB}

We have, so far, studied {\it exact} dualities for the Majorana system 
with the Hamiltonian \(H_{\sf M}\) of Eq. (\ref{Ham}) when subject to periodic boundary conditions.
We focused on periodic boundary conditions these are pertinent to the theoretical study of TQO.  In this appendix,
we will consider {\it exact} dualities in the presence of {\it open} boundary conditions. In doing so, we will further study 
finite, even quite small,  square lattices.  It is useful to provide a precise description
of these finite dual spin systems as there is a definite possibility that this
Majorana architecture may become realizable in the next few years.
These dualities also allow us to illustrate the flexibility of the
bond algebraic approach to dualities in handling a variety of
boundary conditions {\it exactly}. As in the rest of this paper, the dualities
we obtain are exact unitary equivalences. Thus, these dualities 
may be tested numerically by checking if the energy spectra of
the two dual systems are indeed identical. 

\begin{figure}
\includegraphics[angle=0,width=.7\columnwidth]{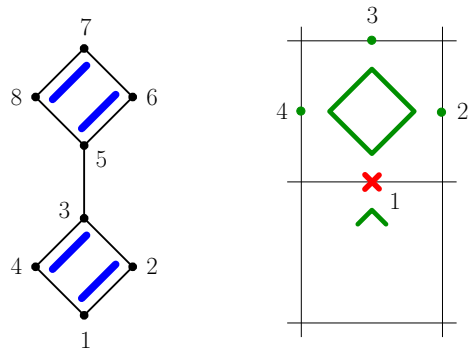}
\caption{The spin dual of two superconducting islands. 
Each island maps to a plaquette interaction of the quantum Ising gauge theory, but such
a mapping would not be compatible with matching dimensions of Hilbert spaces. Hence
one of the lower plaquette is chopped to include only one spin.}
\label{2islands}
\end{figure}

As illustrated in Section \ref{tdn}, the effective Hamiltonian \(H_{\sf M}\) on the
square lattice and in the bulk is dual to the \(\mathds{Z}_2\) lattice gauge theory.
In this appendix, our task  is to find the boundary terms that make the duality exact in the
presence of open boundary conditions. Here we only consider dualities
that preserve the dimension of the Hilbert space of the two theories. We thus follow two
guiding principles: 1) in the bulk, the dual spin theory remains the \(\mathds{Z}_2\) 
lattice gauge theory, and 2) on the boundary, we introduce terms that preserve 
both the bond algebra and the dimension of the Hilbert space.
Let us start with the simplest interacting case, that of two islands (grains) linked by one
Josephson coupling, see Fig. \ref{2islands}. In this case, the Hamiltonian of Eq. (\ref{Ham}) reads 
\begin{equation}\label{2islandsham}
H_{\sf M}=-h c_1c_2c_3c_4- h' c_5c_6c_7c_8-J ic_3c_5.
\end{equation}
This Hamiltonian acts on a Hilbert space of dimension \(\dim\mathcal{H}_{\sf M}=2^{8/2}=2^4\).
Thus, the dual theory must contain four spins and some recognizable gauge interactions. 
The result is
\begin{equation}\label{2dual}
H_{\sf QIG}=-h\sigma^z_1-h'\sigma^z_1\sigma^z_2\sigma^z_3\sigma^z_4-J\sigma^x_1.
\end{equation}
where the single spin \(\sigma^z_1\) in the Hamiltonian stands
for an {\it incomplete plaquette}. One can check that the bond algebra is 
preserved and the two spectra are identical. 
\begin{figure}
\includegraphics[angle=0,width=0.94\columnwidth]{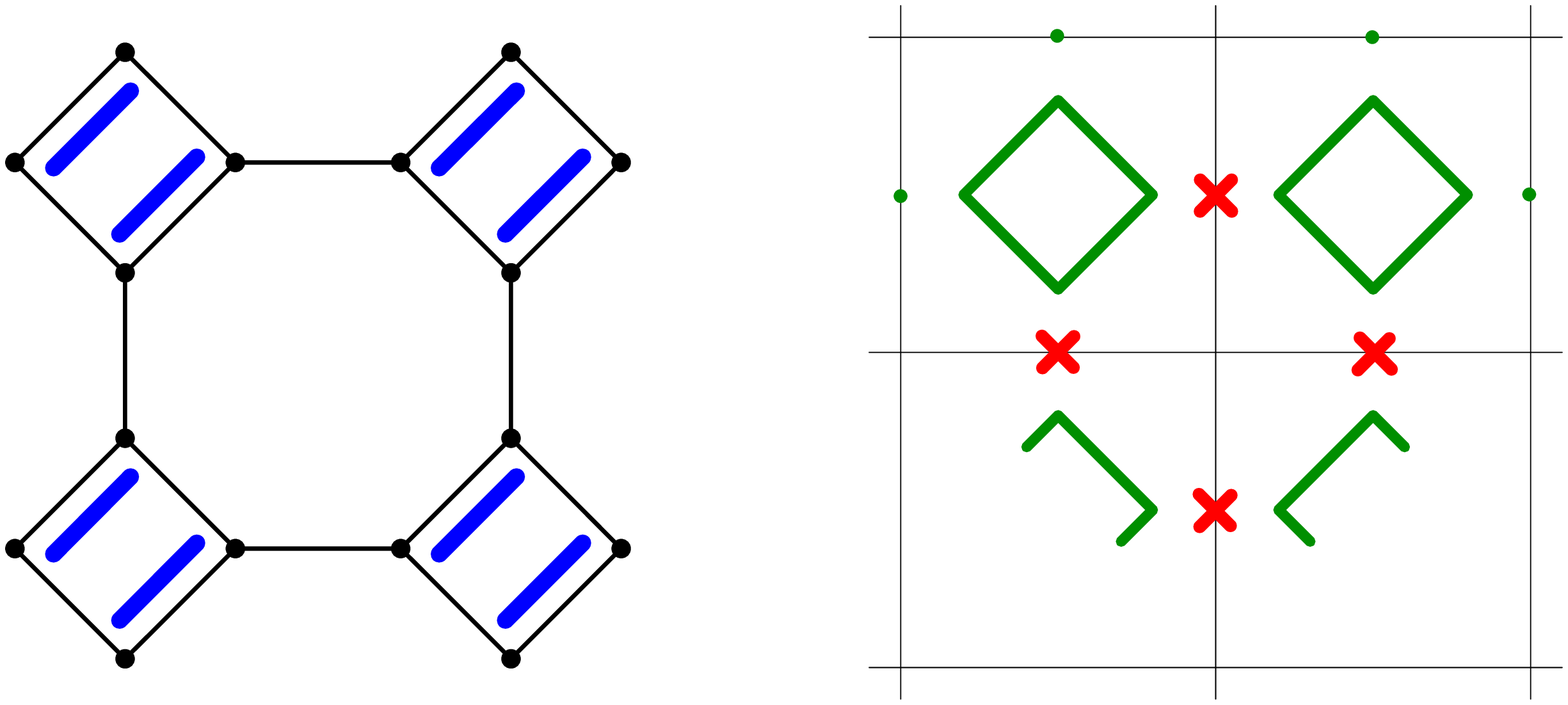}
\caption{The spin dual for a configuration of four islands. The incomplete
plaquettes represent two-spin interactions in the Hamiltonian.}
\label{4islands}
\end{figure}

\begin{figure}
\includegraphics[angle=0,width=.94\columnwidth]{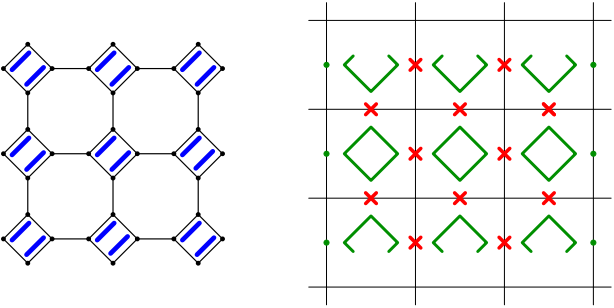}
\caption{The spin dual for nine islands. Incomplete plaquettes
represent three-spin interactions in the spin Hamiltonian, the product of the
three spins \(\sigma^z\) closer to an incomplete green diamond. }
\label{9islands}
\end{figure}

The next interesting case contains four superconducting islands, see 
Fig. \ref{4islands}. In this case, \(\dim\mathcal{H}_{\sf M}=2^{16/2}=2^8\),
and so the dual spin Hamiltonian, described diagrammatically in Fig. \ref{4islands}
contains eight spins, two complete and two incomplete gauge plaquettes. The situation
becomes more regular if we further increase the number of islands. For nine
islands (\(\dim\mathcal{H}_{\sf M}=2^{36/2}=2^{18}\)), the Majorana system maps to eighteen spins, 
three complete, and six incomplete plaquettes
on the first and last row of the spin model. One can generalize this picture
to \(L^2\) islands. Then the dual  \(\mathds{Z}_2\) quantum Ising gauge theory
will be represented by a scaled version of the right panel of 
Fig. \ref{9islands}, with \(2L^2\) spins, and \(2L\) incomplete plaquettes
(the product of only three spins \(\sigma^z\)). The latter incomplete 
plaquettes are equal split between the top and bottom rows,
i.e.,  \(L\) incomplete plaquettes are placed on the top row
and \(L\) are situated on the bottom row.

{\it Notice that there is no natural guiding principle to find the dual theory
by a Jordan-Wigner mapping. The bond-algebraic method is the natural 
approach and can be tested 
numerically on finite lattices. }

\section{Fermionization of $S=1/2$ spin models in arbitrary dimensions} 
\label{appA}

Although not pertinent to our direct models of study 
(those of Eq. (\ref{Ham}) and their exact duals), we briefly review and discuss, 
for the sake of completeness and general perspective, dualities
of related quantum  spin $S=1/2$ systems. 
General bilinear spin Hamiltonians can be expressed 
as a quartic 
form in Majorana fermion operators. The general nature of this
mapping is well known and has been applied to other spin systems with several 
twists. Simply put, we can write each spin operator 
as a quadratic form in Majorana fermions. In the case of general
{\em two-component} spin systems that we discuss now,
the relevant Pauli algebra is given by the following on-site $(\r$) constraints
\begin{eqnarray}
(\sigma^{x}_{\r})^{2} = (\sigma^{z}_{\r})^{2} =1, \ 
\{ \sigma^{x}_{\r}, \sigma^{z}_{\r}\}  =0,
\label{ons}
\end{eqnarray}
and trivial off-site $(\r \neq \r')$ relations,
\begin{eqnarray}
[\sigma^{x}_{\r}, \sigma^{z}_{\r'}]=0.
\label{offs}
\end{eqnarray}
A dual Majorana form may be  
easily derived as follows. We consider a dual Majorana system in which at each 
lattice site $\r$,
there is a grain with three relevant Majorana modes. We label
the three relevant Majorana modes (out of
any larger number of modes on each grain) by $\{c_{\r,a}\}_{a=1}^3$. 
As can be readily seen by invoking Eq. (\ref{majorana_algebra}), a representation 
that trivially preserves the algebraic relations of Eqs. (\ref{ons}, \ref{offs}) is
given by 
\begin{eqnarray}
\label{mapf}
\sigma^{x}_{\r}  \leftrightarrow i c_{\r 1} c_{\r2}, \ 
\sigma^{z}_{\r} \leftrightarrow i c_{\r 1} c_{\r 3}.
\end{eqnarray}
Equation (\ref{mapf}) is a variant of a 
well known
mapping applicable to three component spins  (as well as, trivially, spins with 
any smaller number of components). \cite{Biswas,known}  Equation (\ref{mapf}) may also be viewed 
as a two-component version of the mapping employed by Kitaev. \cite{Kitaev06} The Hilbert space spanned 
by an $S=1/2$ spin system on a lattice/network having
$N$ sites is $\dim{\cal{H}}_{\sf spin} = 2^{N}$. 
By contrast, the Hilbert space of a general Majorana system with $\{ m_{\r}\}$ 
Majorana modes $(m_{\r} \ge 3$)  at sites  $\{\r\}$ is given by 
$\dim {\cal{H}}_{\sf M} 
= 2^{\sum_{\r} m_{\r}/2}$. 
Thus, in this duality the Hilbert space is not preserved: each individual energy level 
of the spin system becomes $2^{(\sum_{\r} m_{\r}/2) - N}$ fold degenerate. 
Similarly, one-component  systems (e.g., those involving only $\{\sigma_{\r}^{x}\}$) 
can be mapped onto a granular system with two Majorana modes per site.
 If there are two Majorana modes at each 
site $\r$ then such a mapping will preserve the Hilbert space size. 

For completeness, we now turn to specific spin systems related to those that we 
discussed in the main part of our article.
In Section \ref{xmcd}, we illustrated that the Majorana system
of Eq. (\ref{Ham}) (and all of its duals that we earlier discussed in the text)
can be mapped onto the Xu-Moore model \cite{XM} on the checkerboard lattice.
Following our general discussion above, it is straightforward to 
provide a Majorana dual to the Xu-Moore model on the square lattice, Eq. (\ref{xme}). 
On the square lattice,
the orbital compass model (OCM) and the Xu-Moore model of Eq. (\ref{xme}) are dual
to one another. \cite{bondprl,ADP,NF}  We will assume the square lattice to define 
the $xz$ plane. The anisotropic square lattice OCM \cite{NF,brink} is 
given by the Hamiltonian
\begin{eqnarray}
\label{ocm+}
H_{\sf OCM} = -  \sum_{\r} (J_{x;\r} \sigma^{x}_{\r} \sigma^{x}_{\r + \i} + 
J_{z; \r} \sigma^{z}_{\r}   \sigma^{z}_{\r + \j}).
\end{eqnarray}
In Eq. (\ref{ocm+}), we generalized the usual compass model Hamiltonian by allowing 
the couplings $\{J_{x,z} \}$ to vary locally with the location
of the horizontal and vertical links of the square lattice (given by $(\r,\r + 
{\bm e}_{1,2})$ respectively).  By plugging Eq. (\ref{mapf}) into Eq. (\ref{ocm+}),
we can rewrite this (as well as other general two-component spin bilinears) as a 
quartic form in the Majorana fermions. 
\begin{figure}[h]
\includegraphics[width=.8\columnwidth]{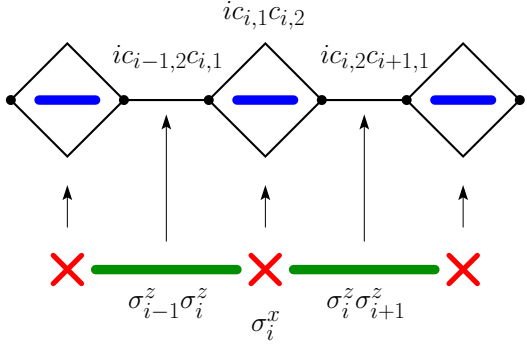}
\caption{The transverse-field Ising model can be simulated by an architecture of
nanowires with one wire per superconducting island.}
\label{ising_majorana}
\end{figure}

It may generally be feasible to use our formalism to 
simulate quantum spin models in terms of Majorana networks. Consider, for 
example,  the simulation of a transverse-field Ising chain 
\begin{equation}
H_{\sf I}= -
\sum_{i=1}^{N-1} J_i\sigma^z_{i}\sigma^z_{{i+1}}-\sum_{i=1}^{N}h_i\sigma^x_{i}.
\end{equation}
with $N$ spins and open boundary conditions. In this case, it may be possible to use linear 
arrays with one nanowire per island to simulate this model and study, for instance, 
the dynamics of its quantum phase transition.
The Hamiltonian $H_{\sf I}$ maps to the Majorana network
\begin{equation}
H_{\sf M}=-i\sum_{i=1}^{N-1} J_i c_{i,2}c_{i+1,1}- i \sum_{i=1}^{N}h_ic_{i,1}c_{i,2} ,
\end{equation}
after the following duality mapping
\begin{equation}
 \sigma^z_i \sigma^z_{i+1}\mapsto i c_{i,2}c_{i+1,1} \ , \  
\sigma^x_i \mapsto i c_{i,1}c_{i,2},
\end{equation}
see Fig. \ref{ising_majorana}.


\begin{thebibliography}{28}
\bibitem{combinatorics}
On each grain, the $z_{\r}$ non-intersecting nanowires link one half of
the nanowire endpoints to  the remaining half; there are
$(2z_{\r})!/(2^{z_{\r}} z_{\r}!)$ distinct ways for  different pairings
of the vertices. These non-intersecting nanowires can be  placed in any
way on a surface of the bulk superconducting grain. For instance, in 
the square and triangular lattices, the regular arrangement of nanowires
shown  in Figs. \ref{l12}, and \ref{l13} is only one among many others.
\bibitem{irreps}
It us useful at this point to recall some basic algebraic facts about 
Majorana fermions. Let us label the Majorana operators simply
as \(c_i, i=1,\cdots,N\). Then, 
\begin{equation}\label{msimpl}
\{c_i,c_j\}=2\delta_{i,j},\ \ \ \ c_i^\dagger=c_i.
\end{equation} 
In general,
\begin{equation}\label{g5}
(c_1\cdots c_N)^2=(-1)^{N(N-1)/2}\mathds{1},
\end{equation}
so that the eigenvalues of \(c_1\cdots c_N\) are \(\pm1\) or \(\pm i\), depending on \(N\).
If \(N\) is even, up to equivalence, there is only one irreducible 
representation (irrep) for these relations. It acts on a Hilbert space of dimension
\(\dim \mathcal{H}_{\sf M}(N \ \mbox{\sf even})=2^{N/2}\), and can be described 
in terms of Pauli matrices as
\begin{equation}
c_i=\left\{
\begin{array}{ll}
\sigma^x_i(\sigma^z_{i-1}\cdots\sigma^z_1),\ \ & i=1,\cdots,\frac{N}{2},\\
\sigma^y_{i-\frac{N}{2}}(\sigma^z_{i-\frac{N}{2}-1}\cdots\sigma^z_1),\ \ & i
=1+\frac{N}{2},\cdots,N.
\end{array}\right.
\end{equation}
If \(N\) is odd,
\begin{equation}
[c_1\cdots c_N, c_i]=0.
\end{equation}
So in view of Eq. \eqref{g5}, the irreps of Eq. \eqref{msimpl} 
are characterized by an  extra constraint,
\begin{equation}
c_1\cdots c_N=\alpha \mathds{1},\ \ \ \ (\mbox{irrep for \(N\) odd})
\end{equation}
with \(\alpha=\pm 1,\pm i\) depending on \(N\) and the particular irrep.
An explicit irrep is afforded by
\begin{eqnarray}
&&c_i=\\
&&\left \{
\begin{array}{ll}
\sigma^x_i(\sigma^z_{i-1}\cdots\sigma^z_1),\ & i=1,\cdots,\frac{N-1}{2},\\
\sigma^y_{i-\frac{N-1}{2}}(\sigma^z_{i-\frac{N-1}{2}-1}\cdots\sigma^z_1),\ 
& i=\frac{N+1}{2},\cdots,N-1,\\
\sigma^z_1\cdots\sigma^z_{\frac{N-1}{2}},\ & i=N.
\end{array}\right.
\end{eqnarray}
Its dimension is \(\dim \mathcal{H}_{\sf M}(N \ \mbox{odd})=2^{(N-1)/2}\).
\bibitem{topology}
In these and other general systems, not all \(d=1\) gauge-like
symmetries are  independent. If the Majorana system has a trivial
homology (such as that of an infinite plane or a sphere),  no  \(d=1\)
gauge-like symmetries appear: all symmetries involving ${\cal O}(L^{1})$
sites can be  expressed in terms of a product of the local symmetries
and thus are not fundamental. By contrast, for a $D=2$ dimensional
system placed on a torus there are two closed toric cycles $\ell$
independent of the local symmetries. This explains the findings of Ref.
\onlinecite{Terhal} that the Majorana theory has TQO. Some
manifestations of this phenomenon were noticed in numerical simulations
of compact QED \cite{jersak}, though not recognized as a signal of the
presence or absence of TQO. In numerous theories with plaquette and/or
link interactions, the Euler-Lhuillier formula 
\begin{eqnarray}
V -E +F = 2(1-g)
\end{eqnarray}
relating the genus number $g$ of the manifold on which the system is
embedded to  the number of local faces ($F$), edges ($E$), and vertices
($V$) affords us with a knowledge of the number of independent ($d=1$)
symmetry operators (loops around independent cycles) that the system may
have that cannot be written in terms of local operators.  This and
related aspects have been discussed elsewhere \cite{TQO}.
\end{thebibliography}

\end{document}